\documentclass[aps, 
preprint,
eqsecnum,
a4paper,
prd]{revtex4-2}
\usepackage{graphicx}
\usepackage{caption}
\usepackage{subcaption}
\usepackage{amsfonts}
\usepackage{amssymb}
\usepackage{amsbsy}
\usepackage{amsmath}
\usepackage{mathrsfs}
\usepackage{bigints}
\usepackage{tensor}
\usepackage{latexsym}
\usepackage{natbib}
\usepackage{bm}
\usepackage{color}
\usepackage{xcolor}
\usepackage{wasysym}
\usepackage{hyperref}
\usepackage{physics}
\usepackage{cancel}
\usepackage{float}
\usepackage{bm}
\usepackage{hyperref}
\usepackage{appendix}

\usepackage[dvipsnames]{xcolor}

\begin{document}
\sloppy

\title{ Spin-torsion interaction and geodesic bending}

\author{Sagar Kumar Maity}
\email{{{sagar.physics1729@bose.res.in}}}
\author{Amitabha Lahiri}
\email{amitabha@bose.res.in}

\affiliation{S. N. Bose National Centre for Basic Sciences,\\
JD Block, Sector 3, Salt Lake, WB 700106, INDIA.}


\date{\today}

\begin{abstract}
The intrinsic spin of fermions induces torsion in spacetime, leading to an effective four-fermion interaction. {This affects} the bending of null and timelike geodesics inside a star with a spherically symmetric distribution of gravitationally dense fermionic matter of constant density. Our analysis shows an additional torsional contribution to the 
geodesic deflection, alongside the conventional curvature-induced effect. This {change depends} on the fermion number density, coupling constants of the different species of fermions with different chiralities, and the temperature of the matter distribution. We have shown that the torsion-induced corrections to null geodesic bending remains very small for both low- and high-mass white dwarfs, whereas significantly larger effects may arise in more compact astrophysical objects such as neutron stars. 
\end{abstract}

\maketitle
\newpage

\section{Introduction}
Massive bodies deform the spacetime around them -- even the trajectory of light bends, a famously confirmed prediction of general relativity (GR)~\cite{Dyson:1920cwa}. Gravitational bending of light provides important information on masses of galaxies \cite{1963SPhD....8..119K, Liebes:1964zz}, the nature of dark matter and dark energy~\cite{Benabed:2003pb, Simpson:2004rz, Finelli:2006iz, Narasimha:1994he, 2010RSPTA.368..967E}, and the large scale structure of the universe~\cite{Schneider:1992bmb, Refsdal:1993kf, Narayan:1996ba, Mellier:1998pk, Bartelmann:1999yn}. 
Gravitational lensing also provides a way to test modifications and externsions of GR~\cite{Knox:2005rg, Keeton:2005jd, 2010ApJ...708..750S, Schmidt:2008hc, Schimd:2004nq, Acquaviva:2004fv, Ovgun:2020gjz}.

In this paper, we apply this idea to Einstein–Cartan–Sciama–Kibble (ECSK) theory~\cite{Cartan:1924yea, Kibble:1961ba, Sciama:1964wt} of a particular type. ECSK theory extends GR by introducing torsion, which arises naturally from the inclusion of fermionic matter. A torsion field generated solely by fermions can be eliminated from the action, leaving behind an effective four-fermion interaction which affects the propagation of fermions through matter~\cite{Hehl:1976kj, Hehl:1971qi, Kerlick:1975tr, finkelstein1960spacetime, Gasperini:2013cru, Chakrabarty:2019cau}. As we shall see below, for a certain type of coupling between torsion and fermion spin, this ``geometrical four-fermion interaction'' has intriguing implications in the collapse of dense configurations~\cite{Choudhury:2024vzx} and also in the early universe~\cite{Dolan:2009ni, Bambi:2014uua, Alexander:2008vt}. However, the torsion tensor is still fully antisymmetric in this scenario, which ensures that its contribution vanishes when contracted with the symmetric combination of velocity vectors in the autoparallel equation, so we recover the usual geodesic equation of GR. Nevertheless, torsion does not completely decouple from the dynamics,  since the effective four-fermion interaction terms which result from the elimination of torsion modify the stress tensor, thus altering the metric coefficients and thereby affecting null and timelike geodesics through the modified geometry.

On the other hand, even when the torsion arises purely from fermions, it can gain a symmetric part if the gravitational action is modified by what is called the Holst term~\cite{Holst:1995pc, Ashtekar:2004eh, Fatibene:2007ce, Fatibene:2007ch}. This term simplifies the algebra of constraints of GR and provides a starting point for the loop quantum gravity (LQG) program~\cite{Ashtekar:1987zz, Rovelli:2004tv, Rovelli:2014ssa}. This modification does not change classical field equations in the absence of torsion. However, when fermions are added via the first-order formalism as in ECSK gravity, the torsion generated by them lets the Holst term modify field equations. In particular, the Barbero-Immirzi parameter, which is the coefficient of the Holst term in the action, can become observable in principle, even if quantum aspects of the gravitational field are not taken into account~\cite{Perez:2005pm, Freidel:2005sn, Mercuri:2006um, Bojowald:2007nu, Kazmierczak:2008iw, Alexandrov:2008iy}. As the torsion need not be fully antisymmetric in presence of the Holst term, geodesics will be modified by torsion. Thus there should be an additional contribution  to gravitational lensing from the spin-torsion coupling in quantum gravity.

The main objective of this paper is to study the effect of the spin-torsion interaction on gravitational lensing. 
Since this interaction is significant only in regions of high fermion density, we need to study geodesics in the interior region of a dense object. We have modeled such an object by a static and spherically symmetric star composed of fermions at constant density. The exterior spacetime is described by the Schwarzschild solution. Then we derive the formula for the gravitational deflection of both null and timelike geodesics for the interior region of the star incorporating the affects of spin-torsion coupling. Previous studies of geodesic deflection within the interior of a spherical mass distribution with constant matter density, but without the spin-torsion interaction, have been reported in the literature~\cite{stuchlik2001null, Escribano:2001ew, Ohanian:1973kj, 1971NCimB...6..225L, 1979ApJ...230..249L, Kling:1998is, Poplawski:2024azm}. {For instance, the Sun has been proposed as a gravitational lens for neutrinos propagating through its interior~\cite{gerver1988focusing, Demkov:2000nq, Patla:2007ju}}, but without considering the spins of the  fermions. In our calculation, the energy-momentum tensor of the background matter is modified by the spin-torsion interaction.

The paper is organized as follows. In Sec.~\ref{fermions} we show how the dynamics of fermions on curved spacetime gives rise to torsion, whose generic couplings to the fermions are chiral and non-universal. In the presence of a Holst term the torsion modifies the geodesic equations. The torsion itself is non-dynamical and leads to a four-fermion interaction, which modifies the energy-momentum tensor at finite temperature. In Sec.~\ref{statphys} we calculate $T_{\mu \nu}$ to leading order for a distribution of fermions in neutron stars and white dwarf stars. The geodesic equation inside a star is modified -- we calculate that for null and timelike geodesics in Sec.~\ref{sec:geodesics}. For the purpose of comparison, we analyze null geodesic bending in the absence of spin–torsion coupling in Sec.~\ref{sec:results with no coupling}. We then include the spin–torsion interaction and study null bending for white dwarfs in Sec.~\ref{sec:results for WD} and for neutron stars in Sec.~\ref{sec:results for NS}. Finally, we summarize our results and discuss their implications in Sec.~\ref{Conclusion}. 
\section{Fermions in Curved Space-time}\label{fermions}
Fermions in curved spacetime are most conveniently dealt with in the ECSK formalism, which  starts with tetrads and a spin connection~\cite{Cartan:1924yea, Kibble:1961ba, Sciama:1964wt, Hehl:1976kj, Gasperini:2013cru, Mielke:2017nwt, baez1994gauge, Peldan:1993hi}. The Dirac matrices $\gamma_a$ are defined on an “internal" flat space, isomorphic to the tangent space at each point. The vectors and tensors in the internal space are related to spacetime objects by the vierbein or tetrad fields $e^\mu_a$, a set of four contravariant vector fields satisfying the orthonormality condition
$g_{\mu\nu} e^\mu_{a} e^\nu_{b} = \eta_{ab}\,, 
\quad \eta_{ab} e^a_\mu e^b_\nu = g_{\mu\nu}\,,$
where $g_{\mu\nu}$ is the spacetime metric and $\eta_{ab}$ is the Minkowski metric $\eta_{ab} = \text{diag}(-1, +1, +1, +1)$. 
Spacetime and internal indices are raised and lowered with the spacetime and Minkowski metrics respectively. We have also defined the ``co-tetrad"  fields $e_\mu^a$ as the inverse of the tetrad fields, 
$\label{tetrads}
e^\mu_a e_\nu^a = \delta^\mu_\nu\,, \quad e^\mu_a e^b_\mu = \delta^b_a\,.
$ 
Then gravity is described by a first order action in terms of the spin connection $A_\mu^{ab}$ and tetrads $e_\mu^a$\,,
\begin{equation}
	S[e, A, \psi]= \frac{1}{2\kappa}\int|e|\, d^4 x\, e_a^\mu e_b^\nu F_{\mu \nu}^{a b}[A] + \frac{i}{2} \int |e|\, d^4 x\, \Big[ \big(\Bar{\psi} \gamma^c e^\mu_c D_\mu \psi\big) - \big(\Bar{\psi} \gamma^c e^\mu_c D_\mu \psi\big)^\dagger
	+ 2 m \Bar{\psi} \psi\Big]. \label{actiontorsin}
\end{equation}
Here $|e|$ is the determinant of the co-tetrad $e_\mu^a$, $F^{ab}_{\mu\nu} = \partial_{[\mu} A_{\nu]}^{ab} + A_{[\mu}^{ac} A_{\nu] c}{}^b\,$ is the curvature of the spin connection $A$\,, and $D_\mu$ is the covariant derivative of a spinor $\psi$\,, defined as
\begin{equation}
	D_\mu \equiv \partial_\mu - \frac{i}{4} A_\mu^{ab} \sigma_{ab}\,,
\end{equation}
where $\sigma_{ab} = \frac{i}{2} [\gamma_a, \gamma_b]_-\,.$

The condition that the connection is tetrad-compatible is $\nabla_\mu e_\lambda^a=0$, sometimes referred to as the tetrad postulate. Using this, we can write the components of the affine connection as
\begin{equation}
	\Gamma_{\mu \nu}^\lambda=e_a^\lambda \partial_\mu e_\nu^a+ A_{\mu b}^a e_\nu^b e_a^\lambda\,. \label{affinewithspin}
\end{equation}  
The spin connection can be written as the sum of two parts, 
\begin{equation}
	A_\mu^{ab} = \omega_\mu^{ab} [e] + \Lambda_\mu^{ab}\,, \label{spin-connection}
\end{equation}
where $\omega_\mu^{ab}$ is the torsion-free spin connection and $\Lambda_\mu^{ab}$ is an independent field, commonly referred to as the contorsion in the literature. We can write $\omega_\mu^{ab}$  fully in terms of the tetrads, co-tetrads, and their derivatives, as
\begin{equation}
	\omega_\mu^{ab} = \frac{1}{2} e^{\nu a}\left(\partial_\mu e_\nu^b-\partial_\nu e_\mu^b\right)-\frac{1}{2} e^{\nu b}\left(\partial_\mu e_\nu^a-\partial_\nu e_\mu^a\right)-\frac{1}{2} e^{\rho a} e^{\sigma b}\left(\partial_\rho e_{\sigma c}-\partial_\sigma e_{\rho c}\right) e_\mu^c\,. \label{torsionfreespin}
\end{equation} 
If this $\omega$ is put in place of the spin connection $A$ in Eq.~\eqref{affinewithspin}, we recover the symmetric Christoffel symbols for the torsion-free Levi-Civita connection. The torsion can be expressed as $\Lambda_{\mu b}^a e_\nu^b e_a^\lambda\,.$
The terms containing the derivatives of $\Lambda$ contribute to a total divergence. 
The remaining $\Lambda$-dependent terms in the action~\eqref{actiontorsin} are  
\begin{equation}
	   \int |e| d^4 x \Big( \frac{1}{2\kappa} \Lambda_{[\mu}^{ac} \Lambda_{\nu]}^{db} \eta_{cd} e^\mu_a e^\nu_b + \frac{1}{8} \Lambda_\mu^{ab} e^\mu_c \Bar{\psi} [\gamma^c, \sigma_{ab}]_+ \psi \Big)\,.
\end{equation} 
Thus $\Lambda$ is a non-propagating or auxiliary field.

The invariance of this action under local Lorentz transformations means that $\Lambda$ transforms homogeneously under them. In particular, the last term of the above action is invariant on its own. It follows that $\Lambda$ can be treated independently from the  universal general relativistic interaction --  different species of fermions may couple to the contorsion field  $\Lambda$ with different coupling constants without violating local or global Lorentz symmetries. Furthermore,  the left- and right-chiral fermions transform independently under local and global Lorentz transformations, so terms like $\Lambda_\mu^{ab} e^\mu_c \Bar{\psi}_L [\gamma^c, \sigma_{ab}]_+ \psi_L$ or $\Lambda_\mu^{ab} e^\mu_c \Bar{\psi}_R [\gamma^c, \sigma_{ab}]_+ \psi_R$ are separately invariant. Therefore $\Lambda$ may couple to left- and right-chiral fermions with distinct coupling constants. The interaction can thus be written as~\cite{Chakrabarty:2019cau}
\begin{align}\label{Lambda-coupling}
        \frac{1}{8} \Lambda_\mu^{ab} e^\mu_c \sum_f 
    \left(\lambda_{fL}\Bar{\psi}_{fL} [\gamma^c, \sigma_{ab}]_+ \psi_{fL} + \lambda_{fR}\Bar{\psi}_{fR} [\gamma^c, \sigma_{ab}]_+ \psi_{fR}\right)\,\notag\\
   \equiv \frac{1}{4}  \Lambda_\mu^{ab} e^\mu_c \epsilon^{cabd} \sum_f  \big( \lambda_{fL} \Bar{\psi}_{fL} \gamma_d \psi_{fL} + \lambda_{fR} \Bar{\psi}_{fR} \gamma_d \psi_{fR} \big)\,,
\end{align}
where the sum runs over all types of fermions.

For the sake of convenience, we reorgarnise the fermions in terms of vector and axial currents, so that the solution for the contorsion tensor is
\begin{equation}
    \Lambda_\mu^{ab} = \frac{\kappa}{4} \epsilon^{cabd} e_{\mu c} \sum_f \big( \lambda_{fV} \Bar{\psi}_f \gamma_d \psi_f + \lambda_{fA} \Bar{\psi}_f \gamma_d\gamma^5 \psi_f \big)\,, \label{contorsiongen}
\end{equation}
for appropriately defined $\lambda_{V,A}$\,. The torsion tensor resulting from~\eqref{contorsiongen} is completely antisymmetric, which ensures that it does not appear in the geodesic equation. As a result, it does not modify the geodesic motion and hence has no effect on gravitational lensing. However, as we shall see below, if the action is modified by the Holst term, the torsion tensor is no longer fully antisymmetric and geodesics may be affected.

The Holst term is crucial to Loop Quantum Gravity. In the absence of torsion it is a total divergence, but when torsion is present it can have observable consequences~\cite{Freidel:2005sn, Perez:2005pm, Benedetti:2011nd}. The gravitational action including the Holst term is written as 
\begin{equation}\label{Holst.1}
    S_{\text{Holst}} = \frac{1}{2\kappa} \int d^4 x |e| e^\mu_a e^\nu_b \Big( F_{\mu\nu}^{ab} - \frac{1}{2\gamma} \epsilon^{ab}{}_{cd} F_{\mu\nu}^{cd} \Big)\,,
\end{equation}
where $\gamma$ is a dimensionless constant known as the Barbero–Immirzi (BI) parameter~\cite{BarberoG:1994eia, Immirzi:1996di}. For vanishing torsion, the Holst modification preserves the Einstein equations but alters the canonical structure of the theory. When spinor fields are included, they can generate torsion, so the Holst action modifies the dynamics of spacetime due to the spin-torsion coupling.

Adding fermions to the gravitational action of Eq.~\eqref{Holst.1}, we get
\begin{equation}
	\begin{split}
		S [e, A, \psi] = \frac{1}{2\kappa}  \int |e| d^4 x &P^{ab}{}_{cd} F^{cd}_{\mu\nu} [A] e^\mu_a e^\nu_b\\ &+ \frac{i}{2} \sum_f\int |e| d^4 x \Big[ \big(\Bar{\psi}_f \gamma^c e^\mu_c D_\mu \psi_f\big) - \big(\Bar{\psi}_f \gamma^c e^\mu_c D_\mu \psi_f\big)^\dagger
		+ 2 m \Bar{\psi}_f \psi_f\Big]\,, \label{ECSKHolstaction}
	\end{split}
\end{equation}
where we have written $P^{ab}{}_{cd} = \frac{1}{2} \big(\delta_c^{[a} \delta_d^{b]} - \frac{1}{\gamma} \epsilon^{ab}{}_{cd} \big)$\,.

In the presence of fermions, we write the spin connection as in (\ref{spin-connection}) and couple the torsion to the fermions with different coupling constants for different particle types and chiralities as in Eq.~(\ref{Lambda-coupling}). Then the generic form of the action is
\begin{align}
	S [e, \Lambda, \psi] =& \frac{1}{2\kappa}  \int |e| d^4 x\, P^{ab}{}_{cd} \Tilde{F}^{cd}_{\mu\nu} [e, \omega] e^\mu_a e^\nu_b + \frac{1}{2\kappa} \int |e| d^4 x\, P^{ab}{}_{cd} \Lambda_{[\mu}^{cp} \Lambda_{\nu]}^{qd} \eta_{pq} e^\mu_a e^\nu_b \notag\\ 
        &+ \sum_f \frac{i}{2} \int |e| d^4 x \left[ \big(\Bar{\psi}_f \gamma^c e^\mu_c \Tilde{D}_\mu \psi_f\big) - \big(\Bar{\psi}_f \gamma^c e^\mu_c \Tilde{D}_\mu \psi_f\big)^\dagger
		+ 2 m \Bar{\psi}_f \psi_f\right] \notag \\
        &+ \sum_f\frac{1}{8} \int |e| d^4 x\, \Lambda_\mu^{ab} e^\mu_c \left[ \lambda_{fL}\Bar{\psi}_{fL} [\gamma^c, \sigma_{ab}]_+ \psi_{fL} + \lambda_{fR}\Bar{\psi}_{fR }[\gamma^c, \sigma_{ab}]_+ \psi_{fR} \right].  \label{totalaction2}
\end{align}  
As before, $\Lambda$ is an auxiliary or non-propagating field. The equation of motion of $\Lambda$ obtained from the action~\eqref{totalaction2} is algebraic,  
\begin{equation}
	\Lambda_{\mu ab} = \frac{\kappa}{4} \frac{\gamma}{\gamma^2 + 1} e_\mu^c \sum_f\Big[ \gamma \epsilon_{cabd} \left(\lambda_V \Bar{\psi}_f\gamma^d\psi_f+\lambda_A \Bar{\psi}_f\gamma^d\gamma_5\psi_f\right) - 2 \eta_{c[a} \left(\lambda_V \Bar{\psi}_f\gamma_{b]}\psi_f+\lambda_A \Bar{\psi}_f\gamma_{b]}\gamma_5\psi_f\right)\Big]\,,\label{spincontorsionholst}
\end{equation} 
where we have used the identity $[\gamma_c, \sigma_{ab}]_+ = 2\epsilon_{abcd}\gamma^d\gamma^5$ and also defined $\lambda_{V,A}=\frac{1}{2}\left(\lambda_R\pm \lambda_L\right)$\,. We can see that the contorsion tensor is not entirely antisymmetric, so as a result we expect to get a contribution of the spin-torsion coupling in the  gravitational lensing, as we will demonstrate below.

We can now insert the solution of $\Lambda$ back into the
action~\eqref{totalaction2}, thus eliminating $\Lambda$ from the theory in favour of a four-fermion interaction term in the action,
\begin{equation}
    \frac{3\kappa}{8} \frac{\gamma^2}{\gamma^2+1} \sum_f\int |e| d^4 x\, \Big[ \lambda_V \Bar{\psi}_f\gamma^d\psi_f+\lambda_A \Bar{\psi}_f\gamma^d\gamma_5\psi_f\Big]^2\,.
\end{equation} 
\subsection{Geodesic Equations}
Since the contorsion is no longer completely antisymmetric in the presence of the Holst term, the geodesic equation will be modified. The resulting deviation from geodesic motion can be interpreted as an effective torsion-induced force, which may have observable consequences, for example in gravitational lensing.
To see this, we consider a parambetrized curve $x^\mu = x^\mu(\lambda)$, with affine parameter $\lambda$ and tangent vector $u^\mu = \frac{d x^\mu}{d \lambda}$. The presence of torsion leads to corrections to the geodesic equation,
\begin{equation}
	u^\alpha \tilde{\nabla}_\alpha u^\mu = - \big(\Lambda_\alpha^{ab} e_{\beta a} e^\mu_b \big) u^\alpha u^\beta. \label{geodesicequ}
\end{equation}
The torsional part which encodes the effects of torsion due to the spin density of matter, while the torsion-free part governs the behavior of spacetime in the absence of spin sources, leading to the usual description of gravity in GR. The covariant derivative $\tilde{\nabla}$ is described by the Christoffel symbols $\tilde{\Gamma}^\mu_{\alpha \beta}$.

For point particles, their trajectories \eqref{geodesicequ} can be influenced by torsion-induced modifications to the spacetime geometry. From Eq.~\eqref{spincontorsionholst} and \eqref{geodesicequ} we write,
\begin{equation}
	 u^\mu \tilde{\nabla}_\mu u^\alpha = - \frac{1}{4} \frac{\gamma}{\gamma^2 + 1} \Big[ \big(u^\mu u^\beta\big) e_{\beta a} \langle J^a \rangle - \big(u^\beta u_\beta\big) e^\mu_a \langle J^a \rangle \Big]\,, \label{autoparallelequation}
\end{equation}
where the current $J^a = \lambda_V \Bar{\psi}\gamma^a\psi+\lambda_A \Bar{\psi}\gamma^a\gamma_5\psi$, includes contributions from both the vector and axial-vector currents.

We now consider a general static and spherically symmetric spacetime with the line element
\begin{equation}
	ds^2 = - B(r) dt^2 + A(r) dr^2 + r^2 (d\theta^2 + \sin^2\theta d\phi^2)\,, \label{metric}
\end{equation}
where~\cite{Schwarzschild:1916ae}
\begin{align}
	B(r) &= \Big( 1 - \frac{r_s}{r} \Big), \qquad r \geq r_g \notag \\
	&= \frac{1}{4}\left(3 \sqrt{1-\frac{r_s}{r_g}}-\sqrt{1-\frac{r^2 r_s}{r_g^3}}\right)^2, \qquad r \leq r_g\,,  \label{B(r)} \\
		A(r) &= \Big( 1 - \frac{r_s}{r} \Big)^{-1}, \qquad r \geq r_g   \notag \\
	&= \left(1-\frac{r^2 r_s}{r_g^3}\right)^{-1}, \qquad r \leq r_g\,. \label{A(r)}
\end{align}
This corresponds to a non-rotating star composed of an incompressible fluid with constant density. Here $r_g$ is the physical radius of the star, and $r_s$ is its gravitational radius.

The components of the tetrads ($e_\alpha^a$) for the metric \eqref{metric} are
\begin{equation}
    e_0^0 = \sqrt{B(r)}, \quad e_1^1 = \sqrt{A(r)}, \quad e_2^2 = r, \quad e_3^3 = r\sin \theta\,.
\end{equation}
The geodesic equations (\ref{autoparallelequation}) for the metric (\ref{metric}) are
\begin{align}
        &\ddot{r} + \frac{A'(r)}{2A(r)} \dot{r}^2 - \frac{r}{A(r)} \dot{\theta}^2 - \frac{r \sin^2\theta}{A(r)} \dot{\phi}^2 + \frac{B'(r)}{2A(r)} \dot{t}^2 = - \xi \Big[ \big(u^\rho \langle J_\rho \rangle \big) \dot{r} - \epsilon \frac{1}{\sqrt{A(r)}} \langle J_1 \rangle \Big]\,, \label{geo.reqn} \\
        &\ddot{\theta} + \frac{2}{r} \dot{\theta} \dot{r} - \sin \theta \cos \theta \dot{\phi}^2 = - \xi \Big[ \big(u^\rho \langle J_\rho \rangle \big) \dot{\theta} - \epsilon \frac{1}{r} \langle J_2 \rangle \Big]\,, \label{geo.thetaeqn} \\
        &\ddot{\phi} + \frac{2}{r} \dot{\phi} \dot{r} + 2 \cot \theta \dot{\phi} \dot{\theta} = - \xi \Big[ \big(u^\rho \langle J_\rho \rangle \big) \dot{\phi} - \epsilon \frac{1}{r \sin \theta} \langle J_3 \rangle \Big]\,, \label{geo.phieqn} \\
        &\ddot{t} + \frac{B'(r)}{B(r)} \dot{t} \dot{r} = - \xi \Big[ \big(u^\rho \langle J_\rho \rangle \big) \dot{t} + \epsilon \frac{1}{\sqrt{B(r)}} \langle J_0 \rangle \Big] \label{geo.teqn} \,.
\end{align}
Here we have defined $\xi = \frac{1}{4} \frac{\gamma}{\gamma^2 + 1}$, and $J_\rho = e_\rho^a J_a = e_\rho^a \big(  \lambda_V \Bar{\psi}\gamma_a\psi+\lambda_A \Bar{\psi}\gamma_a\gamma_5\psi \big)\,.$ The condition of the geodesics to be null is $\epsilon = g_{\mu\nu} u^\mu u^\nu = 0$ and for timelike $\epsilon = -1$, where $u^\mu$ is the tangent to the geodesics.

Extremizing the total action with respect to the vierbein $e_\mu^a$\,, we find the Einstein equations,
\begin{equation}
    \Tilde{R}_{\mu\nu} - \frac{1}{2} \Tilde{R} g_{\mu\nu} = \kappa T_{\mu\nu}\,,
\end{equation}
where the energy-momentum tensor $T_{\mu\nu}$ is
\begin{equation}
    T_{\mu\nu} = \frac{1}{4} \sum_f \Big[ \Bar{\psi}_f \gamma_{(\mu} \Tilde{D}_{\nu)} \psi_f +h.c. \Big] + \frac{1}{2} g_{\mu\nu} \Bigg[ \sum_f  \big( \lambda_{fV} \Bar{\psi}_f \gamma_I {\psi}_f + \lambda_{fA} \Bar{\psi}_f \gamma_I \gamma_5 {\psi}_f \big) \Bigg]^2. \label{emtensor}
\end{equation}
Here $\lambda_{fV}$ and $\lambda_{fA}$ have been rescaled by a factor $\sqrt{\frac{3\kappa}{4}\frac{\gamma^2}{\gamma^2 + 1}}$. We will now derive an expression for $\langle T_{\mu\nu}\rangle$ in degenerate fermionic matter, such as in white dwarfs or neutron stars.
\section{Fermionic energy densities in compact stars}\label{statphys}
The properties of the fermionic background depend crucially on the physical conditions inside the star. In neutron stars all the fermions are degenerate, with Fermi temperatures much larger than the actual temperature of the star. In contrast, white dwarf matter consists of degenerate electrons providing the dominant pressure support, while the nucleons behave as non-degenerate fermionic gases. These two astrophysical environments require slightly different treatments when calculating the expectation value. We discuss them separately below.

The number density of a femionic gas is given by
\begin{equation}
    n = \int_{m}^\infty d\epsilon 
    \frac{\mathcal{D}(\epsilon)}{e^{\beta(\epsilon-\mu)}+1}\,, \label{numberdensity}
\end{equation}
with
\begin{equation}
    \mathcal{D}(\epsilon) = \frac{1}{\pi^2} \epsilon \sqrt{\epsilon^2 - m^2}\,, \label{DOS}
\end{equation}
and all calculations are in natural units, $\hbar = 1$, $c=1$.
\subsection{Neutron Star}

{To begin with, let us consider the case of neutron stars, which can be treated as a gas of degenerate neutrons with $\beta (\mu - m) \gg 1$\,. Then the integral in Eq.~\eqref{numberdensity} can be evaluated (see Appendix~\ref{appendix:a}),\\ }
\begin{equation}
    n = \int_{m}^\mu \mathcal{D}(\epsilon) d\epsilon + \frac{\pi^2}{6} (k_B T)^2 \mathcal{D}'(\mu) + \mathcal{O}(T^4)\,. \label{number density}
\end{equation}

For a system with fixed particle number and fixed volume, as is the case for a static stellar configuration, the number density cannot change with temperature. Using Eq.~\eqref{number density} we can therefore write $\mu$ as a function of temperature as 
\begin{equation}
    \mu(T) = \epsilon_F - \frac{\pi^2}{6} (k_B T)^2 \frac{\mathcal{D}'(\epsilon_F)}{\mathcal{D}(\epsilon_F)} + \mathcal{O}(T^4)\,, \label{muexpression2}
\end{equation}
where $\epsilon_F = \mu(0)$ is the Fermi energy. The number density is then
\begin{equation}
   n = \frac{k_F^3}{3\pi^2}\,.
\end{equation}
where $k_F$ is the fermi momentum, with $\epsilon_F^2 = k_F^2 + m^2$. In order to find an expression for $\langle T_{\mu\nu}\rangle$\,, we split the energy-momentum tensor of~\eqref{emtensor} as
\begin{equation}
    T_{\mu\nu} = T_{\mu\nu}^{\text{free}} + T_{\mu\nu}^{\text{int}}\,,
\end{equation}
where
\begin{equation}
    T_{\mu\nu}^{\text{free}} = \frac{1}{4} \sum_f \Big[ \Bar{\psi}_f \gamma_{(\mu} \Tilde{D}_{\nu)} \psi_f +h.c. \Big]\,,
\end{equation}
and
\begin{equation}
    T_{\mu\nu}^{\text{int}} = \frac{1}{2} g_{\mu\nu} \Bigg[ \sum_f  \big( \lambda_{fV} \Bar{\psi}_f \gamma_a {\psi}_f + \lambda_{fA} \Bar{\psi}_f \gamma_a \gamma_5 {\psi}_f \big) \Bigg]^2\,,
    \label{Tmunuint}
\end{equation}
 where the sum runs over all fermion fields in the theory.
We will see later that this object affects the angle of bending through the total mass of the star. Therefore, we evaluate the expectation value of this in a state corresponding to the gas of fermions being considered.

Throughout our analysis, we assume that the torsional four-fermion interaction is sufficiently weak, providing only a small correction to the free fermion dynamics. Let us briefly discuss what is the allowed size of the couplings $\lambda$. Even though this interaction arises from spacetime geometry, the contorsion field $\Lambda$ is completely independent of the tetrad fields (or the metric), and thus also of the Levi-Civita spin connection $\omega$. So there is no reason for $\Lambda^\mu{}_{a b}$ to have the same coupling to fields as $\omega^\mu{}_{a b}$. Furthermore the coupling of $\omega$ to matter is determined by the classical limit of the theory. However there is no classical limit of the spin-torsion interaction, so it is not possible to assign a value to $\lambda$. All we can say is that the scale of $\lambda$ is not much larger than that of the weak interactions, because otherwise its existence would have been observed in precision experiments already.

We note that even when $\omega$ is very small, the four-fermion interaction remains unaffected. Then the fermion fields can be expanded in terms of plane-wave solutions of the Dirac equation in flat space as is usually done for nucleons in neutron stars. We also assume that the all spins are equally probable in the fermion gas. The expectation values are then computed using the standard Fock-space quantization of the Dirac field and by performing a statistical average over the fermionic distribution characterized by the chemical potential $\mu$ and temperature $T$~\cite{Choudhury:2024vzx}.

Then we can calculate the free part of the energy density as
\begin{align}
     \rho^{\text{free}} &= \langle T_{\mu\nu}^{\text{free}} \rangle u^\mu u^\nu = \frac{2}{(2\pi)^3} \sum_f \int d^3 k_f \frac{(k_{f\mu} u^\mu)^2}{k_f^0} \frac{1}{e^{\beta(k_f^0 - \mu_f)}+1}\,, \notag\\
     & = \frac{1}{\pi^2} \sum_f \int_{m_f}^\infty d\epsilon_f   
    \frac{\epsilon_f^2 ({\epsilon_f^2 - m_f^2})^{\frac{1}{2}}}{e^{\beta(\epsilon_f-\mu_f)}+1}\,. \label{rho free part}
\end{align}
In the approximation $\beta (\mu - m) \gg 1$ we can calculate this as (see Appendix \eqref{appendix:a})
\begin{equation}
    \rho^{\text{free}} = \rho^{\text{free}}_0 + \frac{\pi^2}{6} (k_B T)^2 \sum_f \mathcal{D}(\epsilon_{fF}) + \mathcal{O}(T^4)\,, \label{rhofree}
\end{equation}
where $\rho^{\text{free}}_0$ is the energy density at $T=0$,
\begin{equation}
    \rho^{\text{free}}_0 = \sum_f\int_{m_f}^{\epsilon_{fF}} \epsilon \mathcal{D}(\epsilon) d\epsilon = \sum_f\frac{m_f^4}{8\pi^2} \Big[ z_{f}(2z_{f}^2 +1) \sqrt{1+z_{f}^2} - \sinh^{-1} z_{f} \Big]\, ,\label{rho 0 free}
\end{equation}
with $z_{f} = \dfrac{k_{Ff}}{m_f} = \dfrac{(3\pi^2 n_f)^\frac{1}{3}}{m_f}\,.$ The interaction energy density can be written as
\begin{equation}
    \rho^{\text{int}} = \langle T_{\mu\nu}^{\text{int}} \rangle u^\mu u^\nu = -\frac{1}{2} \sum_f \big( \lambda_{fV}^2 - \lambda_{fA}^2 \big) \langle J_f^{Va} J_{fVa} \rangle + \sum_{f\neq f'} \lambda_{fV} \lambda_{f'V} \langle J_f^{Va} \rangle \langle J_{f'Va} \rangle. \label{rhointeraction}
\end{equation}
The expectation values of vector and axial currents are calculated as described in~\cite{Choudhury:2024vzx},
\begin{equation}
    \langle J_{Va} \rangle = \langle \Bar{\Psi} \gamma_a \Psi \rangle = \frac{1}{(2\pi)^3} \int \frac{d^3k}{k^0} k^a \frac{1}{e^{\beta(k^0-\mu)}+1}, \qquad \langle J_{Aa} \rangle = \langle \Bar{\Psi} \gamma_a \gamma_5 \Psi \rangle = 0 \,.
\end{equation}
Using spherical coordinates and exploiting isotropy, we see that all spatial components of the vector current vanish. Thus we find
\begin{equation}
    \langle J_{V0} \rangle = \frac{1}{\pi^2} \int d\epsilon \frac{\epsilon \sqrt{\epsilon^2 - m^2 }}{e^{\beta(\epsilon - \mu)}+1}, \quad \langle \Vec{J}_V \rangle = 0\,.
\end{equation}
Comparing with  Eq.~\eqref{numberdensity}, we see that 
\begin{equation}
    \langle J_{V0} \rangle = n = \frac{k_F^3}{3\pi^2 }\,. \label{fermi momentum and number density}
\end{equation}
The average of the quadratic can be calculated 
using the same method as before,
\begin{align}\label{expectation JVI JVI}
    \langle J_{Va} J^{Va} \rangle = - \frac{m^2}{(2\pi)^6}\int d^3k' &\int d^3k \frac{1}{k'^0 k^0} \frac{1}{e^{\beta(k'^0-\mu)}+1} \frac{1}{e^{\beta(k^0-\mu)}+1} \notag \\  
    &+ \frac{1}{2} \frac{1}{(2\pi)^6} \int d^3k' \int d^3k\frac{k^ak'_a}{k'^0 k^0} \frac{1}{e^{\beta(k'^0-\mu)}+1} \frac{1}{e^{\beta(k^0-\mu)}+1}\,.
\end{align}
which reduces to
\begin{align} \label{vector-current-square}
    \langle J_{Va} J^{Va} \rangle = - \frac{m^2}{4\pi^4} &\int d\epsilon' \int d\epsilon \frac{\sqrt{\epsilon^2 - m^2 }}{e^{\beta(\epsilon - \mu)}+1} \frac{\sqrt{\epsilon'^2 - m^2 }}{e^{\beta(\epsilon' - \mu)}+1}\\ \notag
    &- \frac{1}{8\pi^4} \int d\epsilon' \int d\epsilon \frac{\epsilon \sqrt{\epsilon^2 - m^2 }}{e^{\beta(\epsilon - \mu)}+1} \frac{\epsilon' \sqrt{\epsilon'^2 - m^2 }}{e^{\beta(\epsilon' - \mu)}+1}\,.  
\end{align}
We thus calculate the energy density due to the torsional interaction as (see Appendix \eqref{appendix:a}),
\begin{align} 
    \rho^{\text{int}} =&  \sum_f \big( \lambda_{fV}^2 - \lambda_{fA}^2 \big) \Bigg[ \frac{m_f^4}{8\pi^4} \Bigg\{ \frac{m_f^2}{4} \Bigg(z_f \sqrt{1 + z_f^2 } -  \sinh^{-1} z_f \Bigg)^2 \notag\\ 
    &+ \frac{\pi^2}{6} (k_B T)^2 \frac{z_f}{\sqrt{1 + z_f^2}} \Bigg(z_f \sqrt{1 + z_f^2 } - \sinh^{-1} z_f \Bigg)\Bigg\} 
    {+ \frac{n_f^2}{16}} \Bigg] \notag \\
    &+  \sum_{f\neq f'} \lambda_{fV} \lambda_{f'V} n_f n_{f'}\,,\label{rhointresur}
\end{align}
where $z_{f} = \dfrac{k_{Ff}}{m_f} = \dfrac{(3\pi^2 n_f)^\frac{1}{3}}{m_f}\,$ as before. If the torsion field couples only to left-chiral (or right-chiral) components of fermions, we say that the interaction is maximally chiral. Then $\lambda_V = \pm \lambda_A$ and the interaction energy density takes the form
\begin{equation}
    {\rho^{\text{int}}}  = \sum_{f\neq f'} {\lambda}_{fV} {\lambda}_{f'V} n_f n_{f'} \,.  \label{rhointmaxchiralint}
\end{equation}  

When the interaction is maximally chiral, Eq.~\eqref{rhointmaxchiralint} implies that the interaction energy density scales as $\rho^{\text{int}} \sim n_f n_{f'}$, and, in particular, that no explicit temperature dependent corrections arise in the interaction contribution. A closely related form of the interaction energy density was previously discussed in Ref.~\cite{Narain:2006kx, Kouvaris:2015rea, Ellis:2018bkr, Mukhopadhyay:2015xhs, Barbat:2024yvi}. For a single species of fermions with maximally chiral interaction, there is no contribution to the energy density. For the neutron star, we could work with a degenerate fermionic gas and compute all the relevant quantities up to first order in the temperature correction. For white dwarfs however, the electron gas is degenerate, but the gas of nucleons is nonrelativistic and nondegenerate. We next compute the stress-energy tensor for this mixed gas. 
%
\subsection{White dwarf}
In a white dwarf, the free part of the energy density receives contributions from two physically distinct components, the mass of the protons and neutrons, and the kinetic energy density of the degenerate electron gas. The kinetic energy of the nucleons is small compared to their mass, and the free part of the energy density for nucleons is 
\begin{equation}
    \rho^{\text{free}}_{\text{nucleon}} = n_p m_p + n_n m_n = \frac{A}{Z} n_p m_p\,,\label{free energy density WD}
\end{equation}
where the subscripts $p$ and $n$ refer to proton and neutron respectively, $A$ is the mass number and $Z$ is the atomic number. Since the overall electric charge density of the star is zero, we must have
\begin{equation}
    n_p = n_e\,,
\end{equation}
where $n_e$ is the number density of the electron. The free part of the energy density is approximated as the rest mass energy density of the nucleons, given by Eq.~\eqref{free energy density WD}. Assuming a spherical star of constant density, we can write the total mass of nucleons, ignoring interactions, as
\begin{equation}
    M_0^{\text{nucleon}} = \frac{4\pi}{3} \frac{A}{Z} n_e m_p r_g^3\,. \label{M_0 nucleon and n_e}  
\end{equation}
On the  other hand, the mass associated with the electrons is obtained from Eq.~\eqref{rhofree}, 
\begin{equation}
    M_{0}^{\text{electron}} = \frac{4\pi}{3} \Bigg[\frac{m_e^4}{8\pi^2} \Big\{ z_{e}(2z_{e}^2 +1) \sqrt{1+z_{e}^2} - \sinh^{-1} z_{e} \Big\} + \frac{1}{6} (k_B T)^2 z_e \sqrt{1 + z_e^2}\Bigg] r_g^3\,. \label{M_0 electron and n_e}
\end{equation}
This is small compared to the nucleon rest-mass energy density. The total mass due to the free part of the energy density is 
\begin{equation}
    M_0 = M_0^{\text{nucleon}} + M_{0}^{\text{electron}}\,. \label{M_0 and n_e}
\end{equation}

In a white dwarf, the nucleons are non-degenerate and non-relativistic, i.e., $\epsilon_f = m_f + \frac{k_f^2}{2m} + \mathcal{O}(k_f^4)$. Therefore from Eq.~\eqref{numberdensity},
\begin{equation}
    \int_{m_f}^\infty d\epsilon_f \frac{\sqrt{\epsilon_f^2 - m_f^2 }}{e^{\beta(\epsilon_f - \mu_f)}+1} = \pi^2 \frac{n_f}{m_f}\,.
\end{equation}
The expectation value of the square of the vector current (Eq.~\eqref{vector-current-square}) for nucleons is 
\begin{equation}
    \langle J_{fVa} J_f^{Va} \rangle = - \frac{m_f^2}{4\pi^4} \Big[ \pi^2 \frac{n_f}{m_f}\Big]^2 - \frac{1}{8\pi^4} \Big[ \pi^2 n_f \Big]^2 = - \frac{3}{8} n_f^2\,.
\end{equation}
The electrons in white dwarfs are strongly degenerate, and therefore the interaction energy density is given by Eq.~\eqref{rhointresur}, {adapted for electrons.}

Hence the total interaction energy density \eqref{rhointeraction} for white dwarf including the contribution from electrons and nucleons is  
\begin{align} \label{int energy density WD}
    \rho^{\text{int}} =& \big( \lambda_{eV}^2 - \lambda_{eA}^2 \big) \Bigg[ \frac{m_e^4}{8\pi^4} \Bigg\{ \frac{m_e^2}{4} \Bigg(z_e \sqrt{1 + z_e^2 } -  \sinh^{-1} z_e \Bigg)^2 \notag\\ 
    &+ \frac{\pi^2}{6} (k_B T)^2 \frac{z_e}{\sqrt{1 + z_e^2}} \Bigg(z_e \sqrt{1 + z_e^2 } - \sinh^{-1} z_e \Bigg)\Bigg\} 
    {+ \frac{n_e^2}{16}} \Bigg] \notag \\
    &+\frac{3}{16} \sum_{f\in (n, p)} \big( \lambda_{fV}^2 - \lambda_{fA}^2 \big) n_f^2 +  \sum_{\substack{f \neq f' \\ f,f' \in \{n,p,e\}}} \lambda_{fV} \lambda_{f'V} n_f n_{f'}\,.
\end{align}
We will now see how this interaction modifies the geodesic equations and thus the total bending angle experienced by a test particle in the interior of a star. 

\section{Bending of Geodesics}\label{sec:geodesics}

\subsection{Bending of Null Geodesics}
In this section, we study the bending of null geodesics in the interior spacetime given in~Eq.~\eqref{metric}. We restrict to configurations where both the source and the observer lie in the equatorial plane ($\theta = \pi/2$), so that the null geodesic remains confined to this plane. The equations of motion corresponding to the $\phi$ and $t -$ components of the null geodesics can be written using Eq.~\eqref{geo.reqn} -- \eqref{geo.teqn},
\begin{align}
    \dv{}{\lambda} \Bigg[\ln \Big( r^2 \dv{\phi}{\lambda} \Big)\Bigg] &= - \xi u^\rho \langle J_\rho\rangle\,, \label{phigeoequ}\\
%
    \dv{}{\lambda} \Bigg[\ln \Big( B(r) \dv{t}{\lambda} \Big)\Bigg] &= - \xi u^\rho \langle J_\rho\rangle\,. \label{tgeoequ}
\end{align}
The quantity $u^\rho \langle J_\rho\rangle$ is constant along the geodesic, i.e., 
\begin{equation}
    u^\alpha \nabla_\alpha \big( u^\rho \langle J_\rho \rangle \big) = 0, \quad u^\rho \langle J_\rho \rangle  = a\, {\rm (constant)}, \label{urhojrho}
\end{equation}
%
From \eqref{phigeoequ} and \eqref{tgeoequ}, we have
\begin{equation}
     \dv{\phi}{\lambda} = \frac{L}{r^2} e^{-a \xi \lambda}, \quad \dv{t}{\lambda} = \frac{E}{B(r)} e^{-a \xi \lambda}. \label{nullphitequ}
\end{equation}
here $L$ and $E$ are constants of integration related to angular momentum and energy per unit mass, respectively.

The geodesic equation for the $r$-coordinate on the equatorial plane $\theta = \pi/2$ can also be written as 
\begin{equation}
    \dv{r}{\lambda} = \sqrt{\frac{1}{A(r)}\Bigg(\frac{E^2}{B(r)} - \frac{L^2}{r^2}\Bigg)} e^{-a \xi\lambda}. \label{nullrequ}
\end{equation}

We can, therefore, write
\begin{equation}
    \dv{\phi}{r} = \frac{d\phi/d\lambda}{dr/d\lambda} = \frac{L}{r^2} \sqrt{\frac{A(r)}{\frac{E^2}{B(r)} - \frac{L^2}{r^2}}}
\end{equation}
 We define the impact parameter $b = L/E$, then
\begin{equation}
    \dv{\phi}{r} = \frac{b}{r^2} \sqrt{\frac{A(r)}{\frac{1}{B(r)} - \frac{b^2}{r^2}}} . \label{dphidrnew}
\end{equation} 
The total deflection angle is given by the exact integral
\begin{equation}
\alpha
=
2\int_{b}^{\infty}
dr \frac{b}{r^2} \sqrt{\frac{A(r)}{\frac{1}{B(r)} - \frac{b^2}{r^2}}}
-\pi,
\label{eq:deflection_exact}
\end{equation}
In the weak-field regime, we write
\begin{equation}
B(r)=1+2\Phi(r), \qquad A(r)=1-2\Phi(r), \qquad |\Phi|\ll 1.
\end{equation}
To linear order in $\Phi$, 
\begin{equation}
\sqrt{A(r)} \simeq 1-\Phi(r), 
\qquad 
\frac{1}{B(r)} \simeq 1-2\Phi(r).
\end{equation}
Expanding Eq.~\eqref{eq:deflection_exact} to first order in $\Phi$ and evaluating the trajectory along the unperturbed straight-line path $r^2=b^2+z^2$, we find that the deflection angle is
\begin{equation}
\alpha
=
2\int_{-\infty}^{\infty}
\frac{b}{r}\frac{d\Phi}{dr}\,dz,
\label{total bending angle in weak field}
\end{equation}
where $z$ is the distance along the unperturbed straight-line trajectory of the particle and $r=\sqrt{b^2+z^2}\,.$
\begin{figure}[ht]
\begin{center}
  \begin{subfigure}[H]{0.30\textwidth}
    \includegraphics[width=\textwidth]{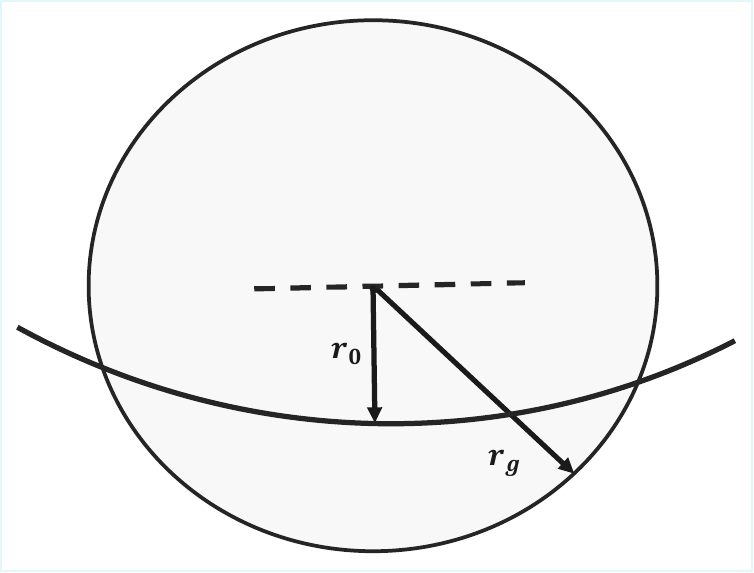}
    \label{fig:f1}
  \end{subfigure}
  \begin{subfigure}[H]{0.24\textwidth}
    \includegraphics[width=\textwidth]{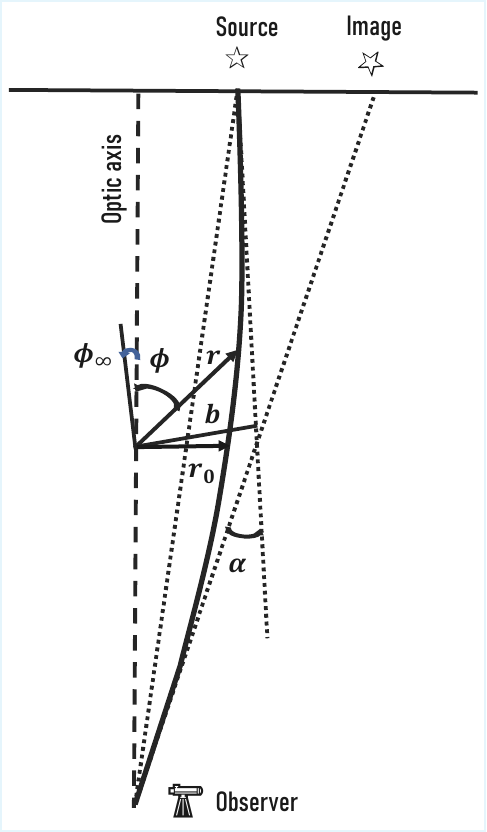}
    \label{fig:f2}
  \end{subfigure}
  \caption{Gravitational lensing by spherically distributed matter with constant density}
\end{center}
\end{figure}
We model the interior as a non-rotating star composed of an incompressible fluid with constant density~\cite{Schwarzschild:1916ae}. The stellar interior is assumed to contain generic matter fields, including both bosonic and fermionic components. Our focus is on the impact of fermionic fields and the associated spin-torsion coupling on the trajectories of test particles within the star. The exterior spacetime is described by the Schwarzschild solution. {For the metric functions in Eq.~\eqref{B(r)} and \eqref{A(r)} we can write}
\begin{equation}
\Phi(r)=
\begin{cases}
-\dfrac{r_s}{4r_g^3}(3r_g^2-r^2), \quad & r<r_g\,,\\ 
-\dfrac{r_s}{2r}, & r\ge r_g\,,
\end{cases}
\end{equation}
where $r_g$ is the physical radius of the star and $r_s$ is its gravitational radius. The interior metric is physical for $r_s < \frac{8}{9} r_g$, otherwise the object must collapse to form a black hole. The trajectory intersects the interior region for $|z|\le z_*$, where $z_*=\sqrt{r_g^2-b^2}$. 

Therefore the interior deflection becomes
\begin{equation}
\Delta\phi_{\rm int}
=
2\int_{-z_*}^{z_*}
\frac{b}{r}\frac{r_s}{2r_g^3}r\,dz
=
\frac{2r_sb}{r_g^3}\sqrt{r_g^2-b^2}\,,
\end{equation}
while the deflection in the exterior is
\begin{equation}
\Delta\phi_{\rm ext}
=
2\int_{z_*}^{\infty}
\frac{b}{r}\frac{r_s}{2r^2}\,dz
=
\frac{2r_s}{b}
\left(
1-\frac{\sqrt{r_g^2-b^2}}{r_g}
\right)\,.
\end{equation}
The total deflection angle~\eqref{total bending angle in weak field} is the sum of the interior and exterior contributions,
\begin{equation}
\alpha
=
\frac{2r_s}{b}
\left(
1-\frac{\sqrt{r_g^2-b^2}}{r_g}
\right)
+
\frac{2r_sb}{r_g^3}\sqrt{r_g^2-b^2}\,.
\end{equation}
In the limit $b \to r_g$, it reduces to the standard Schwarzschild result, while for $b \to 0$ the deflection angle vanishes as expected. {Our result is consistent with the deflection angle obtained in Ref.~\cite{Escribano:2001ew}} for $r_s \ll r_g$.

The gravitational radius is defined as $r_s = 2 G M_{\rm eff}$, with the effective mass for a constant-density configuration is
\begin{equation}
    M_{\rm eff} = M_0 + \Delta M\,, \label{effective mass}
\end{equation}
where $\Delta M = \frac{4\pi}{3} \rho^{\rm int} r_g^3$ arises from the four-fermion interaction term, {while $M_0$ includes the contribution to the stellar mass from the Standard Model interactions and particle masses. } The modified mass function $M_{\rm eff}$ enters directly into the interior metric functions and hence into the null-geodesic equation inside the star. As a result, the interaction energy density $\rho^{\text{int}}$ modifies the bending angle through its contribution to the gravitational potential.

Thus the total bending angle is 
\begin{equation}
\alpha
=
4 G M_{\rm eff}\, \Bigg[\frac{1}{b}
\left(
1-\frac{\sqrt{r_g^2-b^2}}{r_g}
\right)
+
\frac{b}{r_g^3}\sqrt{r_g^2-b^2}\Bigg]\,.\label{total bending angle}
\end{equation}
%

\subsection{Bending of timelike geodesics} 

We now analyze the bending of timelike geodesics for a massive test particle passing through a spherically symmetric stellar interior, focusing on the effects induced by spin–torsion coupling. Timelike trajectories satisfy the normalization condition
\begin{equation}
u^\mu u_\mu = -1 ,
\end{equation}
and the affine parameter is chosen to be the proper time $\tau$.

Starting from the modified geodesic equations \eqref{geo.reqn}--\eqref{geo.teqn}, the equations of motion for the $t$ and $\phi$ coordinates take the form (see appendix~\ref{appendix:a} for details)
\begin{align}
B(r)\,\frac{dt}{d\tau} &= E\,e^{-a\xi\tau}
+ \xi \langle J_0\rangle e^{-a\xi\tau}
\int d\tau' \, e^{a\xi\tau'} B^{1/2}(r) , \nonumber\\
r^2\frac{d\phi}{d\tau} &= L\,e^{-a\xi\tau}
- \xi \langle J_3\rangle e^{-a\xi\tau}
\int d\tau' \, e^{a\xi\tau'} r ,
\label{timelikephitequ}
\end{align}
where $E$ and $L$ denote the conserved energy and angular momentum in the torsion-free limit.

Restricting motion to the equatorial plane, the normalization condition yields
\begin{equation}
- B(r)\Big(\frac{dt}{d\tau}\Big)^2
+ A(r)\Big(\frac{dr}{d\tau}\Big)^2
+ r^2\Big(\frac{d\phi}{d\tau}\Big)^2
= -1 .
\label{timelikecond}
\end{equation}
Using \eqref{timelikephitequ} in \eqref{timelikecond}, we find that the radial equation becomes
\begin{equation}
\frac{dr}{d\tau} = \sqrt{\frac{1}{A(r)}\left[-1+ e^{-2a\xi\tau}\left(\frac{K_1^2(r)}{B(r)}
- \frac{K_2^2(r)}{r^2}\right)\right]},
\label{timelikerequ}
\end{equation}
where we have defined
\begin{align}
K_1(r)
&=
E
+ \xi\langle J_0\rangle
\int_0^{\tau(r)} d\tau' \, e^{a\xi\tau'} B^{1/2}(r)\,,
\\
K_2(r)
&=
L
- \xi\langle J_3\rangle
\int_0^{\tau(r)} d\tau' \, e^{a\xi\tau'} r\,.
\end{align}
We can calculate the orbital equation from this as
\begin{equation}
\frac{d\phi}{dr}
=
\frac{K_2(r)e^{-a\xi\tau}}{r^2}
\sqrt{
\frac{A(r)}
{
-1
+ e^{-2a\xi\tau}
\left(
\frac{K_1^2(r)}{B(r)}
- \frac{K_2^2(r)}{r^2}
\right)
}
}.
\end{equation}

The total deflection angle is defined as
\begin{equation}
\alpha
=
2\big|\phi_\infty - \phi(r_0)\big| - \pi
=
2\Big[
(\Delta\phi)_{\rm ext}
+
(\Delta\phi)_{\rm int}
\Big]
- \pi ,
\end{equation}
where $r_0$ denotes the radius of closest approach and $r_g$ the stellar surface.
The exterior contribution is governed by the Schwarzschild geometry,
\begin{equation}
(\Delta\phi)_{\rm ext}
=
\int_{r_g}^{\infty} dr \,
\frac{L}{r^2}
\sqrt{
\frac{A(r)}
{
-1
+ \left(
\frac{E^2}{B(r)}
- \frac{L^2}{r^2}
\right)
}
}\,,
\end{equation}
while the interior contribution is given by
\begin{equation}
(\Delta\phi)_{\rm int}
=
\int_{r_0}^{r_g} dr \,
\frac{K_2(r)e^{-a\xi\tau(r)}}{r^2}
\sqrt{
\frac{A(r)}
{
-1
+ e^{-2a\xi\tau(r)}
\left(
\frac{K_1^2(r)}{B(r)}
- \frac{K_2^2(r)}{r^2}
\right)
}
}\,.
\label{intbendangtime}
\end{equation}

For a spherically symmetric star the spin-current component $\langle J_3\rangle$ vanishes, implying $K_2=L$. We work in the weak-torsion and weak-field regime, retaining only terms linear in the torsion–matter coupling and assuming
$a^2\ll1$ and $a\langle J_0\rangle\ll1$. This approximation is appropriate for low-compactness objects such as white dwarfs, where torsion effects remain perturbative.

The torsion-free limit corresponds to $a=0$ and $\langle J_0\rangle=0$, yielding
\begin{equation}
K_1 = E , \qquad K_2 = L .
\end{equation}
The interior bending angle in general relativity is therefore
\begin{equation}
(\Delta\phi)_{\rm int}^{(0)}
=
\int_{r_0}^{r_g} dr \,
\frac{L}{r^2}
\sqrt{
\frac{A(r)}
{
-1
+ \left(
\frac{E^2}{B(r)}
- \frac{L^2}{r^2}
\right)
}
}\,.
\label{intbendangtimeGR}
\end{equation}
Including torsion to leading order, we write
\begin{equation}
K_1(r)=E+\delta K_1(r),
\qquad
\delta K_1(r)
=
\xi\langle J_0\rangle
\int_0^{\tau(r)} d\tau' \, B^{1/2}(r),
\end{equation}
and expand Eq.~\eqref{intbendangtime} to linear order in $a\xi\tau$ and $\delta K_1$. The interior deflection angle then becomes
\begin{equation}
(\Delta\phi)_{\rm int}
=
(\Delta\phi)_{\rm int}^{(0)}
+
\int_{r_0}^{r_g} dr \,
\frac{L}{r^2}
\frac{\sqrt{A(r)}}
{\Big[-1+\big(\frac{E^2}{B(r)}-\frac{L^2}{r^2}\big)\Big]^{3/2}}
\left(
a\xi\tau(r)
-
\frac{E\,\delta K_1(r)}{B(r)}
\right).
\label{intbendangtime3}
\end{equation}
The second term in Eq.~\eqref{intbendangtime3} defines the torsional correction,
\begin{equation}
\delta\phi_{\rm torsion}
=
\int_{r_0}^{r_g} dr \,
\frac{L}{r^2}
\frac{\sqrt{A(r)}}
{\Big[-1+\big(\frac{E^2}{B(r)}-\frac{L^2}{r^2}\big)\Big]^{3/2}}
\left(
a\xi\tau(r)
-
\frac{E\,\delta K_1(r)}{B(r)}
\right),
\label{deltaphitorsion}
\end{equation}
so that
\begin{equation}
(\Delta\phi)_{\rm int}
=
(\Delta\phi)_{\rm int}^{(0)}
+
\delta\phi_{\rm torsion}\,.
\end{equation}
This analysis leads us to conclude that torsion contributes explicitly to the bending of timelike geodesics, whereas null geodesics do not acquire any explicit torsional dependence at the level of the
geodesic equation, as shown in Eq.~\eqref{dphidrnew}. For null trajectories, the influence of spin–torsion interactions enters only through modification of the effective mass, which in turn alters the interior metric functions and hence the null geodesic equation inside the star. In timelike geodesic bending, {let us consider only the terms linear in the spin-torsion coupling constants.} At this order, the torsion-induced correction to the mass function is negligible, so the interior metric functions remain effectively unchanged from their general relativistic forms. The torsional contribution to the timelike bending angle arises exclusively from $K_1(r)$ and $K_2(r)$, together with the exponential damping factor appearing in Eq.~\eqref{intbendangtime}.  
\section{Null geodesic without spin-torsion coupling} \label{sec:results with no coupling}
In this section, we study the  bending of null geodesics and present numerical results for the deflection angle for different compact stellar parameters. We start with the geodesic bending in the absence of spin-torsion coupling in the interior of a compact astrophysical objects. {This is the purely metric contribution to geodesic bending, which serves as the reference against which torsion-induced effects will be quantified.}

In the limit $\lambda_V = \lambda_A = 0$, Eq.~\eqref{total bending angle} reduces to the purely gravitational bending angle,
\begin{equation}
\alpha =
4 G M_0\, \Bigg[\frac{1}{b}
\left(
1-\frac{\sqrt{r_g^2-b^2}}{r_g}
\right)
+
\frac{b}{r_g^3}\sqrt{r_g^2-b^2}\Bigg]\,. \label{bending angle alpha}
\end{equation}
\begin{figure}[ht]
  \begin{subfigure}[H]{0.45\textwidth}
    \includegraphics[width=\textwidth]{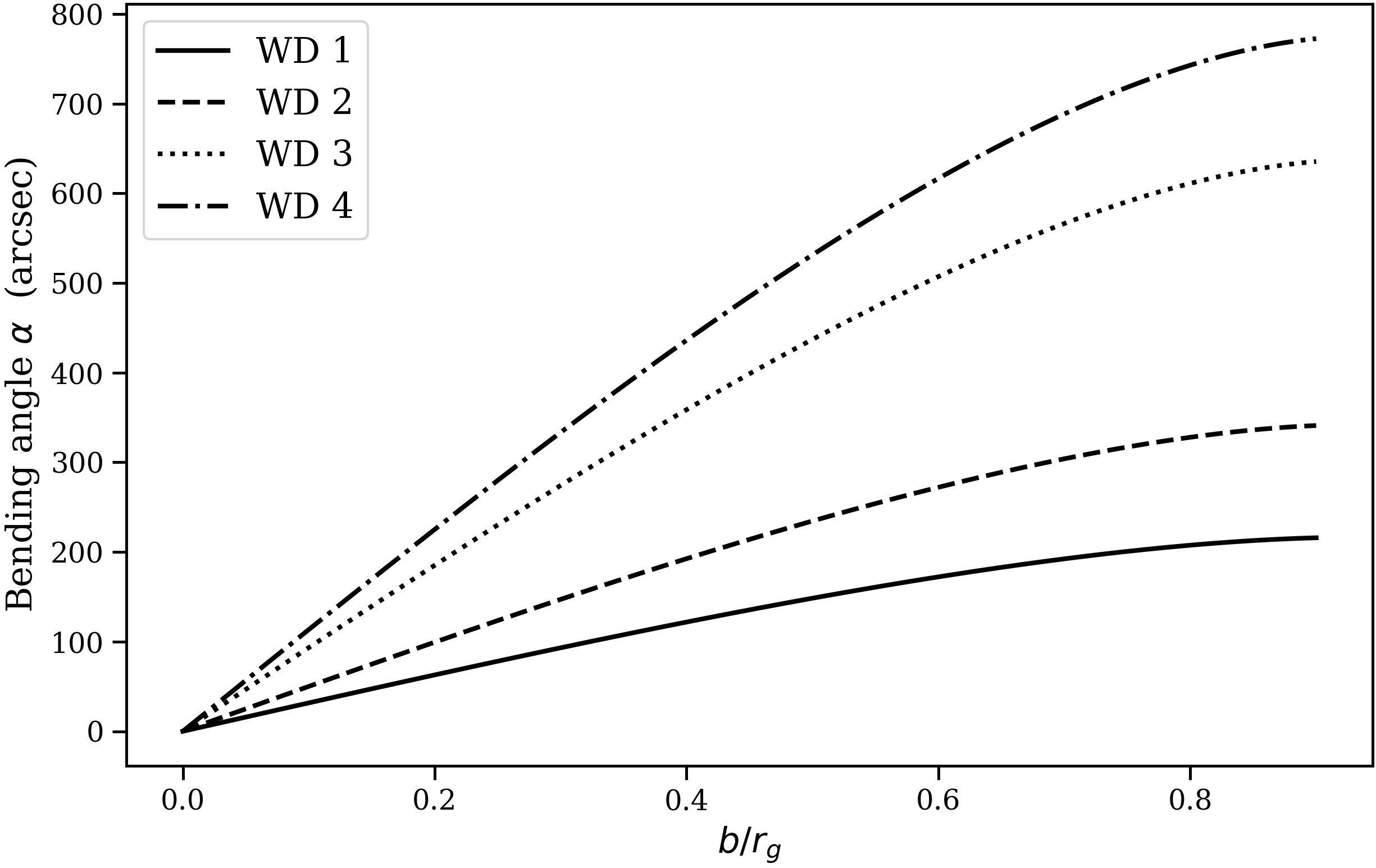}
  \end{subfigure}
  \hspace{0.5cm}
  \begin{subfigure}[H]{0.45\textwidth}
    \includegraphics[width=\textwidth]{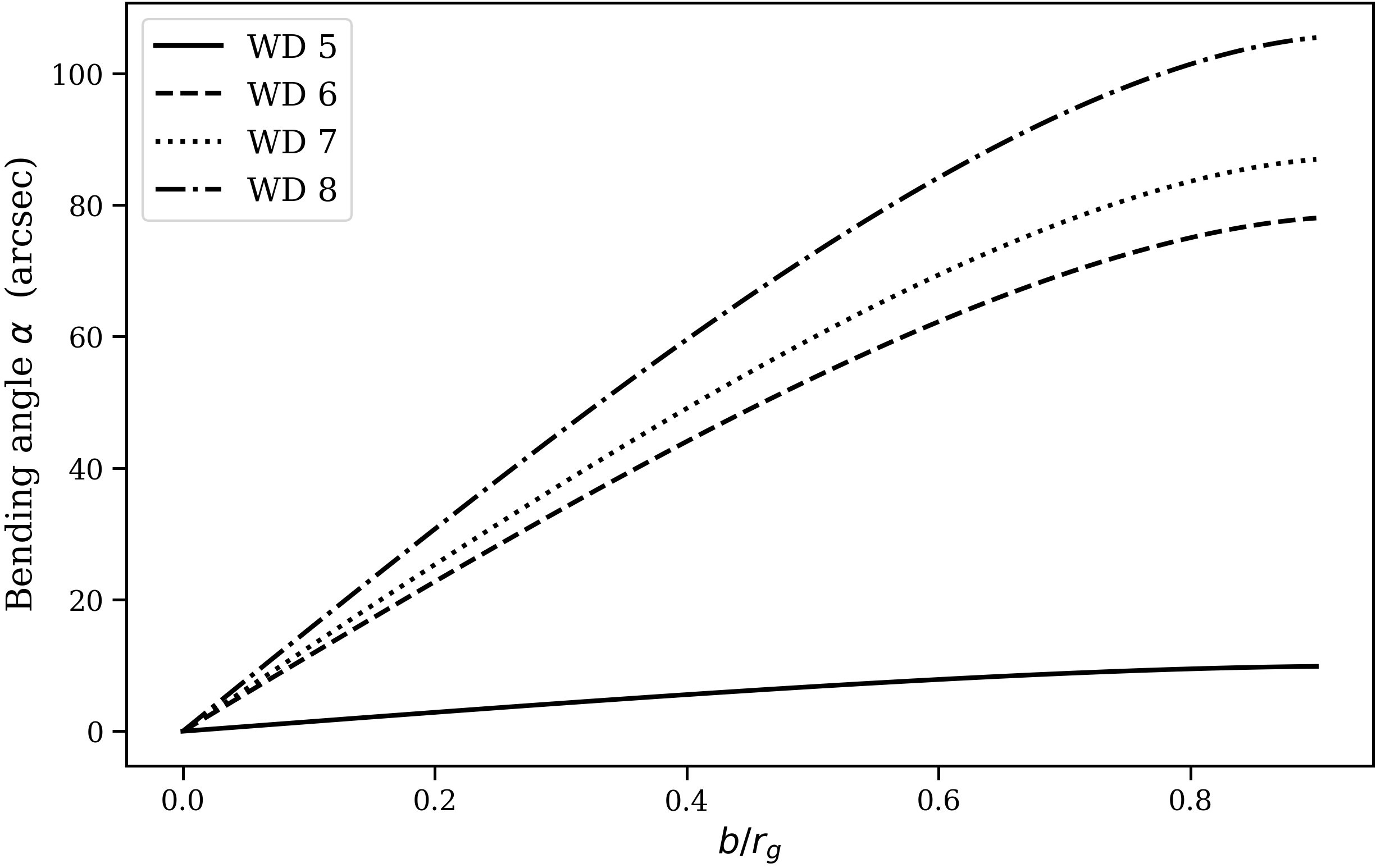}
  \end{subfigure}
  \caption{Total gravitational bending angle $\alpha$ as a function of $b/r_g$ for white dwarfs of different compactness in the absence of four-fermion interaction $(\text{i.e., \,}\lambda_V = 0, \, \lambda_A = 0)$. }
  \label{bendangplots1}
\end{figure}
In Fig.~\ref{bendangplots1} we have plotted this bending angle $\alpha$ as a function of the normalized impact parameter $b/r_g$ for compact stars of various sizes. 
{We see that } the bending angle increases monotonically with $b$ and attains its maximum value at the stellar surface $r_g$. {In the limit $b \to 0$, the deflection angle vanishes due to the spherical symmetry of the mass distribution, ensuring the absence of any net transverse deflection along a radial trajectory}. On the left panel of Fig.~\ref{bendangplots1}, we consider a sequence of compact stellar configurations (WD1-WD4) with parameters chosen to be representative of high-density white dwarf–like objects. Specifically, we take WD1: 
$(M = 1.02\,M_{\odot},\; r_g = 5.8 \times 10^{3}\,\text{km})$, WD2: 
$(M = 1.10\,M_{\odot},\; r_g = 4.0 \times 10^{3}\,\text{km})$, WD3: 
$(M = 1.31\,M_{\odot},\; r_g = 2.6 \times 10^{3}\,\text{km})$, and WD4: 
$(M = 1.35\,M_{\odot},\; r_g = 2.1 \times 10^{3}\,\text{km})$, spanning increasing compactness. 
These parameters correspond to the known white dwarf stars Sirius B, BPM 37093, LHS 4033, and ZTF J1901+1458, respectively.  
As is evident from Fig.~\ref{bendangplots1}, a denser star produces a larger deflection {for a fixed $b/r_g$} due to a stronger gravitational field. On the right panel of Fig.~\ref{bendangplots1}, we consider a complementary sequence of less compact configurations (WD5-WD8), with parameters chosen to represent the low density white dwarfs. In particular, we take WD5: 
$(M = 0.17\,M_{\odot},\; r_g = 21.0 \times 10^{3}\,\text{km})$, WD6: 
$(M = 0.57\,M_{\odot},\; r_g = 9.1 \times 10^{3}\,\text{km})$, WD7: 
$(M = 0.6\,M_{\odot},\; r_g = 8.6 \times 10^{3}\,\text{km})$, and WD8: 
$(M = 0.67\,M_{\odot},\; r_g = 7.9 \times 10^{3}\,\text{km})$. The parameter values correspond to NLTT 11748,  40~Eridani~B, Procyon~B, and Stein~2051~B, respectively. These configurations exhibit systematically smaller bending angles compared to the left panel, reflecting the weaker gravitational fields associated with larger radii and lower masses.   For instance, the most compact configuration considered here (WD4) is representative of extreme high density white dwarf ZTF J1901+1458 and yields a bending angle $\alpha \simeq 800^{\prime\prime}$ at the surface, whereas a low-density extended configuration (WD5), which has size and mass comparable to those of NLTT 11748, produces a much smaller deflection $\alpha \simeq 9.66^{\prime\prime}$.  
\begin{figure}[hbt]
\centering
\includegraphics[width=0.45\textwidth]{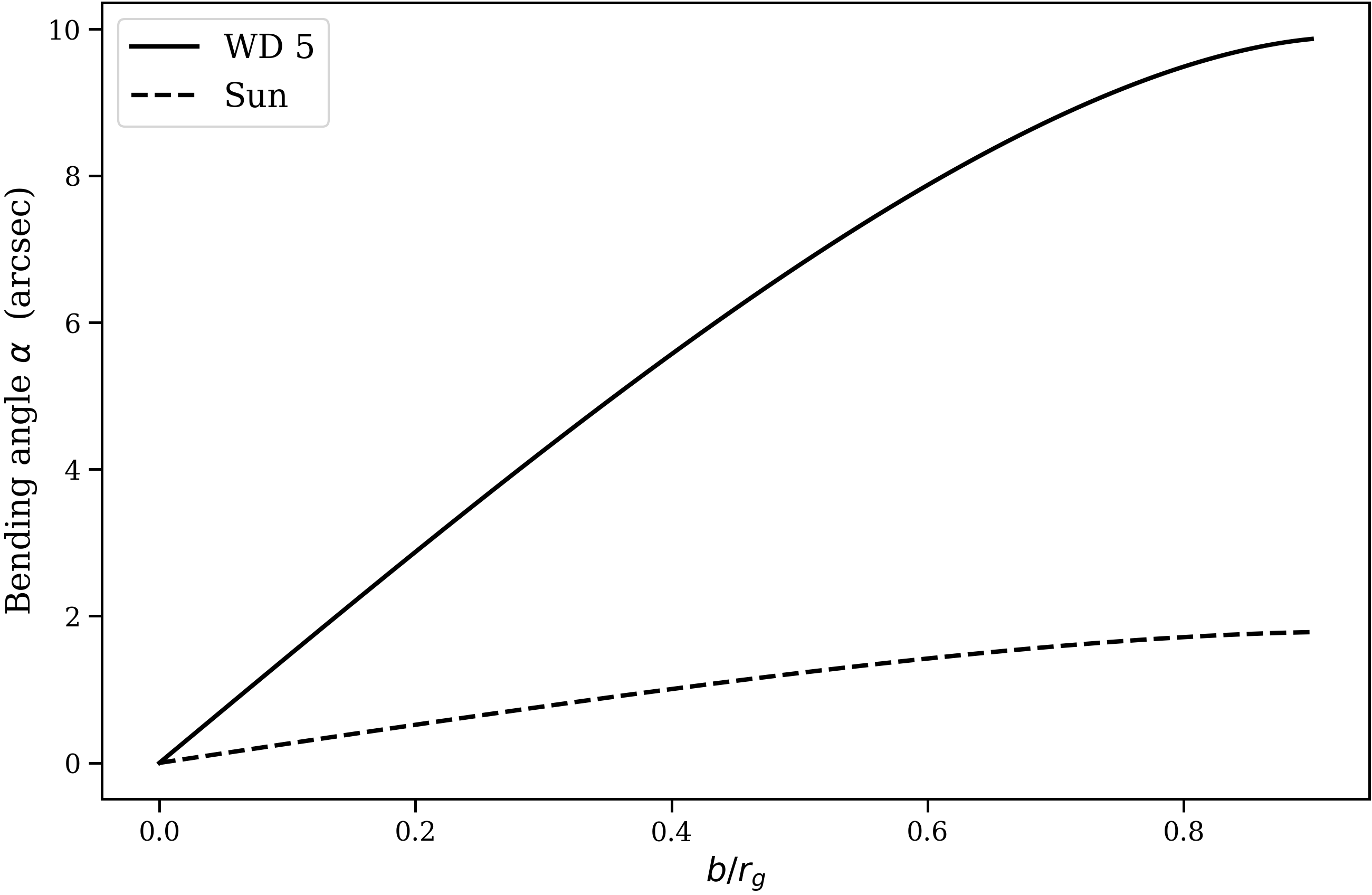}
\caption{Bending angle of null geodesics for WD~5 (NLTT 11748) and the Sun} \label{bendangplots2}
\end{figure}

\begin{figure}[ht]
\centering
\includegraphics[width=0.45\textwidth]{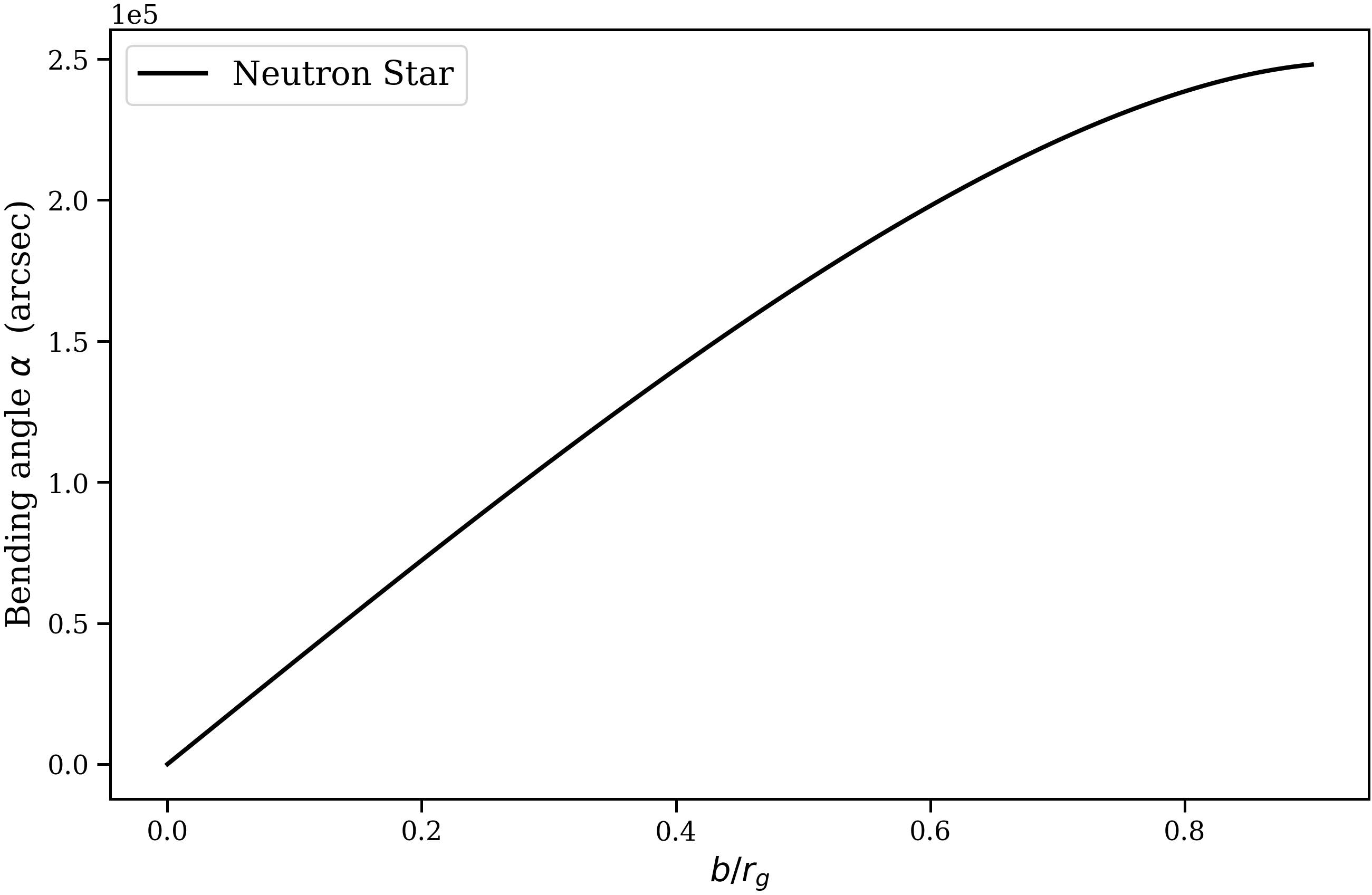}
\caption{Bending angle of null geodesics for neutron star} \label{bendangplots3}
\end{figure}

For reference, we have plotted in Fig.~\ref{bendangplots2} a comparison of the bending angle in Eq.~\eqref{bending angle alpha} for the Sun and a low density white dwarf configuration (WD5). For both of them the bending angle increases with $b/r_g$\,, but the maximum values are different, $1.75''$ for the Sun and $9.66''$ for the white dwarf.  Finally, Fig.~\ref{bendangplots3} shows the behavior of the bending angle in Eq.~\eqref{bending angle alpha} for a neutron star with mass $2\,M_\odot$ and radius $R = 10\,\mathrm{km}$. Due to the extreme compactness of neutron stars, the deflection is substantially enhanced compared to white dwarfs, illustrating the transition toward the strong-field regime of gravitational lensing.

So far, we have considered the gravitational bending of null geodesics in the interior of compact astrophysical objects, such as white dwarfs and neutron stars, in the absence of four-fermion interactions. We have also ignored electroweak interactions, which must be included for actual particles propagating on these geodesics. In the following, we consider the effect of the torsional four-fermion interaction on the bending of geodesics in the interiors of white dwarfs and neutron stars, although we will continue to ignore electroweak interactions.

\section{Null geodesic bending in white dwarfs}\label{sec:results for WD}
Since the strength of the four-fermion term is sensitive to the fermion number density, it is useful to distinguish between white dwarf stars of high and low density of fermions.

\subsection{High density white dwarf}
We first consider a high density white dwarf configuration characterized by a large electron number density $n_e$, which governs both the degeneracy pressure and the strength of the torsion-induced fermionic interactions. 

To model a high density white dwarf approaching the Chandrasekhar regime, we adopt an electron number density 
\begin{equation}
    n_e \sim 10^{37}\,\text{m}^{-3}\,.
\end{equation}
For a representative stellar radius $r_g \simeq 2.6 \times 10^6\,\text{m}$, this density yields a reference mass scale $M_0 \simeq 1.3\, M_\odot$, characteristic of massive white dwarfs near the Chandrasekhar limit, consistent with the properties of LHS 4033. The corresponding electron Fermi momentum, obtained from Eq.~\eqref{fermi momentum and number density}, is
\begin{equation}
   k_{eF} \simeq 1.3 \times 10^{-3}\,\text{GeV}\,, 
\end{equation}
which gives $z_e = \frac{k_{eF}}{m_e} \simeq 2.6\,.$ This indicates that the degenerate electrons in such a massive white dwarf are in the relativistic regime. The effective mass \eqref{effective mass} is therefore
\begin{align} \label{effective mass for high density WD}
    M_{\text{eff}} \simeq& M_0 + \frac{4\pi}{3}r_g^3\Bigg[ \big( \lambda_{eV}^2 - \lambda_{eA}^2 \big) \Bigg\{ \frac{m_e^6}{32\pi^4} \Bigg(z_e \sqrt{1 + z_e^2 } -  \sinh^{-1} z_e \Bigg)^2 \notag\\ 
    &+ \frac{m_e^4}{48\pi^2} (k_B T)^2 \frac{z_e}{\sqrt{1 + z_e^2}} \Bigg(z_e \sqrt{1 + z_e^2 } - \sinh^{-1} z_e \Bigg) 
    {+ \frac{n_e^2}{16}} \Bigg\} \notag \\
    &+\frac{3}{16} \sum_{f\in (n, p)} \big( \lambda_{fV}^2 - \lambda_{fA}^2 \big) n_f^2 +  \sum_{\substack{f \neq f' \\ f,f' \in \{n,p,e\}}} \lambda_{fV} \lambda_{f'V} n_f n_{f'}\Bigg]\,.
\end{align}
For high density white dwarfs, the electrons are strongly degenerate and relativistic, with the three contributions appearing in the effective mass expression at a representative white-dwarf temperature $T = 10^{5}\,\mathrm{K}$. The mass-dependent degeneracy term is $\frac{m_e^6}{32\pi^4} \big(z_e \sqrt{1 + z_e^2 } -  \sinh^{-1} z_e \big)^2 \sim 1.8\times10^{-22}\,\mathrm{GeV}^6$, the thermal correction is $\frac{m_e^4}{48\pi^2} (k_B T)^2 \frac{z_e}{\sqrt{1 + z_e^2}} \big(z_e \sqrt{1 + z_e^2 } - \sinh^{-1} z_e \big) \sim 5.6 \times 10^{-32}\,\mathrm{GeV}^6$, and the density-squared term is $\frac{1}{16}n_e^2 \sim 2\times10^{-21}\,\mathrm{GeV}^6$. The thermal contribution is suppressed by more than ten orders of magnitude relative to the leading term due to $k_B T \ll k_{eF}$ and is therefore entirely negligible for white-dwarf interiors. On the other hand, for the parameters of a white dwarf star, the degeneracy term is comparable with the density square terms.

The bending angle in the interior of the high density white dwarf is
\begin{align} \label{bending angle for high density WD}
    \alpha = & \alpha_{\text{GR}} + \frac{16 \pi G}{3}r_g^3\Bigg[ \big( \lambda_{eV}^2 - \lambda_{eA}^2 \big) \Bigg\{ \frac{m_e^6}{32\pi^4} \Bigg(z_e \sqrt{1 + z_e^2 } -  \sinh^{-1} z_e \Bigg)^2   
    {+ \frac{n_e^2}{16}} \Bigg\}  \notag \\  
    & +\frac{3}{16} \sum_{f\in (n, p)} \big( \lambda_{fV}^2 - \lambda_{fA}^2 \big) n_f^2  + \sum_{\substack{f \neq f' \\ f,f' \in \{n,p,e\}}}\lambda_{fV} \lambda_{f'V} n_f n_{f'}\Bigg] \Bigg[\frac{1}{b}\left(1-\frac{\sqrt{r_g^2-b^2}}{r_g}
\right)+\frac{b}{r_g^3}\sqrt{r_g^2-b^2}\Bigg]\,.
\end{align}
%
%
\begin{figure}[ht]
\includegraphics[width=0.45\textwidth]{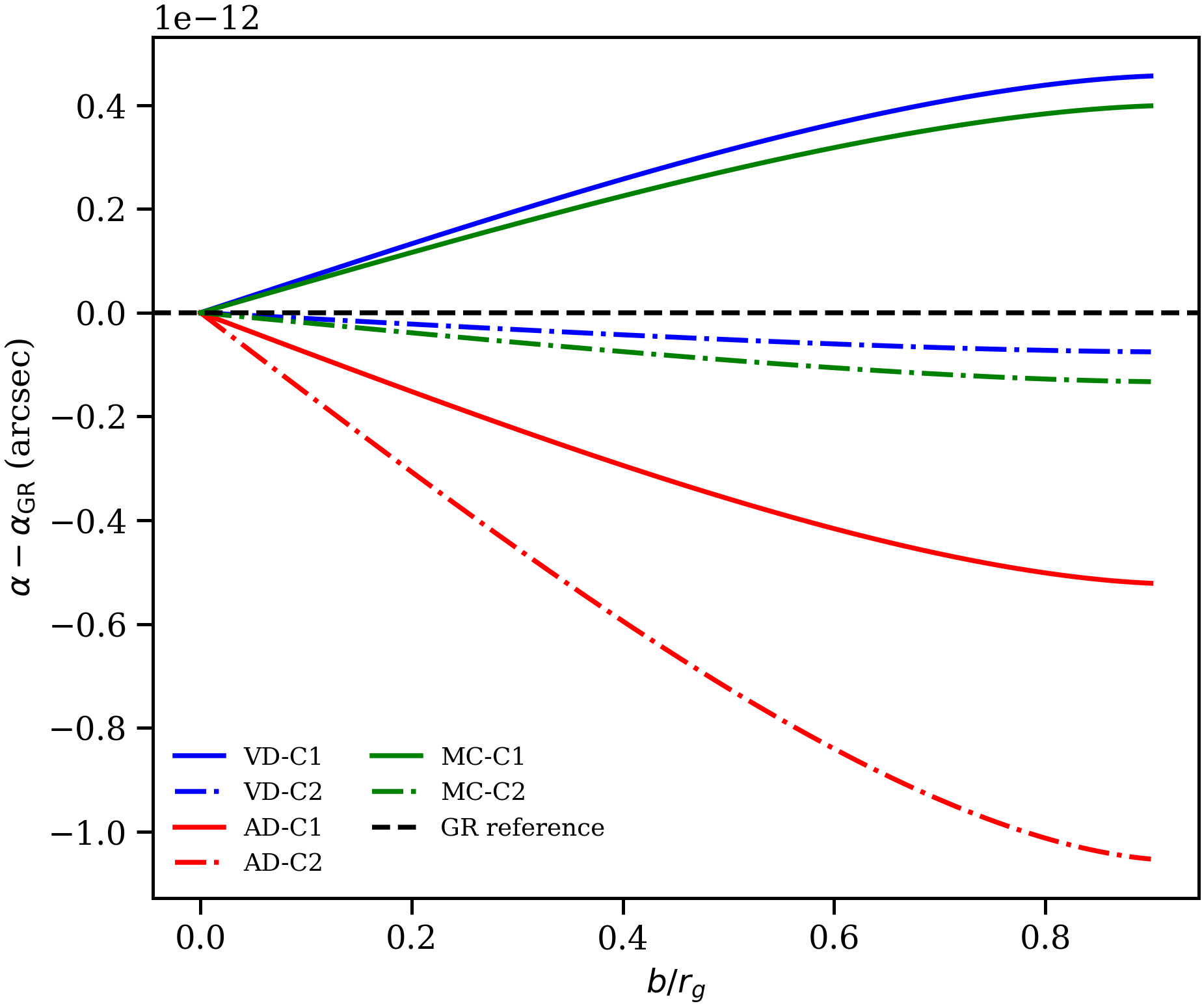}
\caption{
Torsion-induced correction to the bending angle
$\alpha-\alpha_{\rm GR}$ for high density white dwarfs,
as a function of $b/r_g$ for different vector and axial coupling configurations.
}
\label{WD high mass_null_geodesic_bending with impact parameter} 
\end{figure}
%
To study null geodesic deflection in a compact white dwarf, we numerically evaluate Eq.~\eqref{bending angle for high density WD}. Fig.~\ref{WD high mass_null_geodesic_bending with impact parameter} shows the torsion-induced correction, $\alpha-\alpha_{\rm GR}$, as a function of the impact parameter for a high-density white dwarf. Since the four-fermion couplings may have either sign, we consider two representative choices, C1 and C2, shown in Fig.~\ref{WD high mass_null_geodesic_bending with impact parameter}. The couplings are taken to be of order $10^{-3}\,\mathrm{GeV}^{-1}$~\cite{Chakraborty:2024zek}. For C1 (solid curves) we have taken $\lambda_{eV}=\lambda_{pV}=\lambda_{nV}=-10^{-3}\,\mathrm{GeV}^{-1}$, whereas for C2 (dash-dotted curves) we have taken $\lambda_{eV}=-10^{-3}\,\mathrm{GeV}^{-1}$ and $\lambda_{pV}=\lambda_{nV}=10^{-3}\,\mathrm{GeV}^{-1}$; {these two choices capture all physically distinct sign configurations of the vector couplings}. In addition we have taken the axial couplings $\lambda_A$ in three different configurations, vector-dominated: $\lambda_{fV}=4\lambda_{fA}$, axial-dominated: $\lambda_{fA}=4\lambda_{fV}$, and maximally chiral: $\lambda_{fA}=\lambda_{fV}$. The corresponding plots are marked by the labels VD, AD, and MC, respectively. The results show that the torsion-induced correction, $\alpha-\alpha_{\rm GR}$, may become either positive or negative depending on the relative signs and magnitudes of the couplings, corresponding to an enhancement or suppression of the bending angle compared to the torsion-free case (GR). This behavior arises from the modification of the effective gravitational potential by torsional four-fermion interactions. In all coupling configurations considered, $|\alpha-\alpha_{\rm GR}|$ increases monotonically with the normalized impact parameter $b/r_g$ and reaches its maximum value near the stellar surface, $b=r_g$. The resulting correction to the bending angle is of order $\sim10^{-13}\, \mathrm{arcsec}$.  
%
\begin{figure}[ht]
\includegraphics[width=0.5\textwidth]{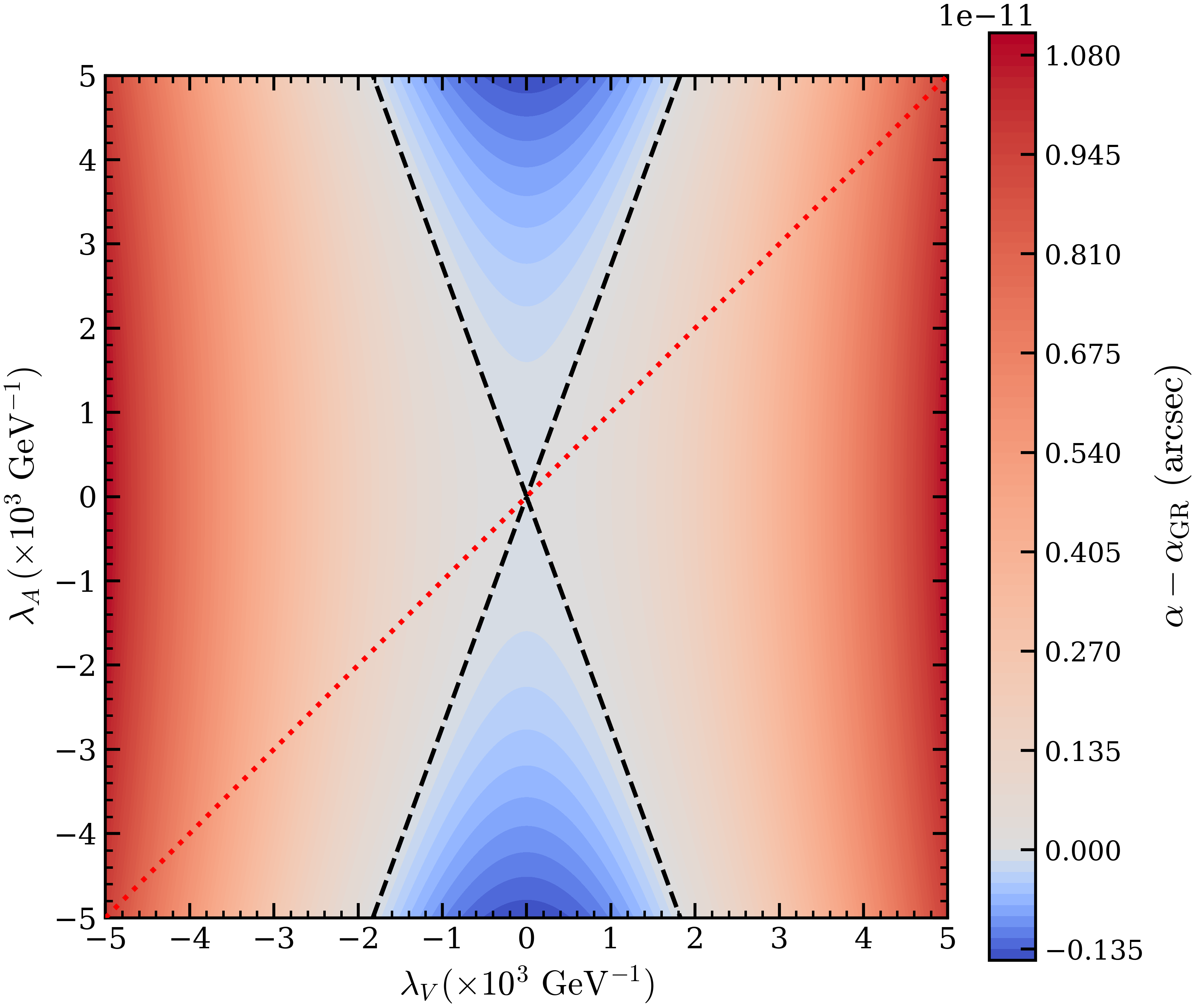}
\caption{
Filled contour plot of the torsion-induced correction to the bending angle $\alpha-\alpha_{\rm GR}$, in the $(\lambda_V,\lambda_A)$ parameter space for a high-density white dwarf with $b/r_g=0.8$. 
}
\label{WD high mass_null_geodesic_bending contour plot} 
\end{figure}
%

Fig.~\ref{WD high mass_null_geodesic_bending contour plot}
shows the torsion-induced correction to the bending angle,
$\alpha-\alpha_{\rm GR}$, in the $(\lambda_V,\lambda_A)$ parameter space for a high-density white dwarf, evaluated at a fixed impact parameter $b/r_g=0.8$. The vector and axial couplings are $\lambda_V,\lambda_A\sim\mathcal{O}(10^{-3})~{\rm GeV}^{-1}$, with $\lambda_{eV}=\lambda_{pV}=\lambda_{nV}\equiv\lambda_V$
and $\lambda_{eA}=\lambda_{pA}=\lambda_{nA}\equiv\lambda_A$. The coupling ranges considered in this plot, $\lvert\lambda_V \rvert,\,\lvert\lambda_A\rvert \lesssim 5\times10^{-3}\ {\rm GeV}^{-1},$ are of the same order as the weak interaction scale set by $\sqrt{G_F} \simeq 3.4\times10^{-3}\ {\rm GeV}^{-1}$. The figure shows that spin--torsion interactions can enhance (red regions) or suppress (blue regions) the gravitational deflection relative to the torsion-free case. The black dashed contour corresponds to $\alpha=\alpha_{\rm GR}$. The positive corrections occur in the vector-dominated regime, while the correction becomes negative for sufficiently large axial coupling, approximately when $|\lambda_A|\gtrsim 2.74\,|\lambda_V|$. The correction remains nonzero along the maximally chiral line $\lambda_V=\lambda_A$ (red dotted line), which is due to the mixed fermionic interaction terms in the effective energy density, and hence in the bending angle \eqref{bending angle for high density WD}. For couplings of order $\mathcal{O} (10^{-3})~\mathrm{GeV}^{-1}$, the resulting modification is small, of order $\sim10^{-12}\,\mathrm{arcsec}$,  while still exhibiting a nontrivial sensitivity to the underlying microscopic fermionic interaction structure.  
\subsection{Low-density white dwarf}
We now discuss the bending of null geodesics in the interior of a low-density white dwarf configuration characterized by a smaller electron number density $n_e$. To model a low-density white dwarf {we set the electron number density to be}
\begin{equation}
    n_e \sim 10^{35}\,\text{m}^{-3}\,.
\end{equation}
For a characteristic stellar radius $r_g \simeq 8 \times 10^6\,\text{m}$, this density corresponds to a reference mass scale $M_0 \simeq 0.6\,M_\odot$ which is typical of low-density white dwarfs such as Procyon B. The corresponding Fermi momentum obtained from Eq.~\eqref{fermi momentum and number density} is $k_{eF} \simeq 2.8 \times 10^{-4}\,\text{GeV}$,
which gives $z_e = \frac{k_{eF}}{m_e} \simeq 0.55$. {Thus these electrons} are at best  mildly relativistic. {Going back to }Eq.~\eqref{int energy density WD} {and expanding in powers of $z_e$, we can write }  
\begin{equation}
   z_e \sqrt{1 + z_e^2 } -  \sinh^{-1} z_e \approx \frac{2z_e^3}{3} - \frac{z_e^5}{5} + \mathcal{O}(z_e^7)\,,
\end{equation}
and therefore the effective mass \eqref{effective mass} for this low-density configuration is  
\begin{align} \label{effective mass for low density WD}
    M_{\text{eff}} \simeq M_0 + \frac{4\pi}{3}r_g^3\Bigg[ &\big( \lambda_{eV}^2 - \lambda_{eA}^2 \big) \Bigg\{     - \frac{m_e^6}{120\pi^4} z_e^8 + \frac{m_e^4}{48\pi^2} (k_B T)^2 \Bigg(\frac{2}{3} z_e^4 - \frac{8}{15} z_e^6 \Bigg) 
    {+ \frac{3n_e^2}{16}} \Bigg\} \notag \\
    &+\frac{3}{16} \sum_{f\in (n, p)} \big( \lambda_{fV}^2 - \lambda_{fA}^2 \big) n_f^2 +  \sum_{\substack{f \neq f' \\ f,f' \in \{n,p,e\}}} \lambda_{fV} \lambda_{f'V} n_f n_{f'}\Bigg]\,.
\end{align}
Evaluating the interaction terms at $T = 10^{5}\,\mathrm{K}$, we find $\tfrac{3}{16} n_e^2 \sim 1.1 \times 10^{-25}\,\mathrm{GeV}^6$, and $\frac{m_e^6}{120\pi^4} z_e^8 \sim 1.3 \times 10^{-26}\,\mathrm{GeV}^6$, whereas the thermal correction is suppressed to $\sim 10^{-32}\,\mathrm{GeV}^6$. The interaction energy density is therefore dominated by the density-squared term, and the mass dependent term, with thermal effects playing no role in the bending-angle correction for low-density white dwarfs. 
The bending angle in the interior of the low density white dwarf is,
\begin{align} \label{bending angle for low density WD}
    \alpha = \alpha_{\text{GR}} +& \frac{16 \pi G}{3}r_g^3\Bigg[ \big( \lambda_{eV}^2 - \lambda_{eA}^2 \big) \Bigg\{     - \frac{m_e^6}{120\pi^4} z_e^8 
    {+ \frac{3n_e^2}{16}} \Bigg\} + \frac{3}{16} \sum_{f\in (n, p)} \big( \lambda_{fV}^2 - \lambda_{fA}^2 \big) n_f^2 \notag \\  
    &+  \sum_{\substack{f \neq f' \\ f,f' \in \{n,p,e\}}} \lambda_{fV} \lambda_{f'V} n_f n_{f'}\Bigg] \Bigg[\frac{1}{b}\left(1-\frac{\sqrt{r_g^2-b^2}}{r_g}
\right)+\frac{b}{r_g^3}\sqrt{r_g^2-b^2}\Bigg]\,.
\end{align}
%
%
\begin{figure}[ht]
\centering
\includegraphics[width=0.45\textwidth]{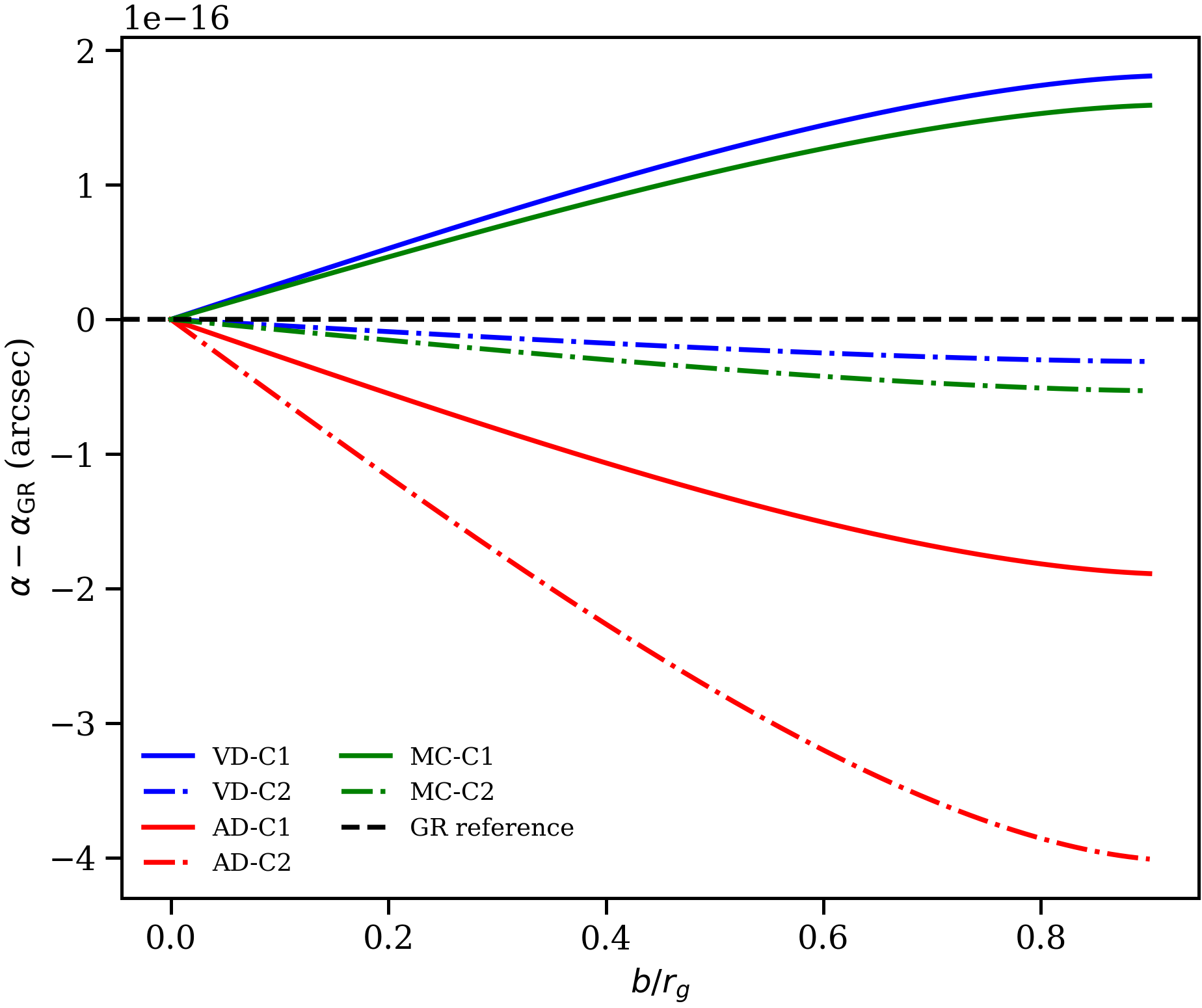} 
\caption{Bending angle correction due to spin-torsion coupling for low density white dwarfs as a function of $b/r_g$ for different vector and axial coupling configurations.} 
\label{WD low mass_null_geodesic_bending with impact parameter}
\end{figure}
%
We compute the bending angle in Eq.~\eqref{bending angle for low density WD} using the effective mass given in Eq.~\eqref{effective mass for low density WD} for a low-density white dwarf. Fig.~\ref{WD low mass_null_geodesic_bending with impact parameter} shows the torsion-induced correction $\alpha-\alpha_{\rm GR}$ as a function of the impact parameter. The same coupling configurations and parameter choices as those used for the high-density white dwarf are adopted here. Depending on the coupling configuration, the correction can be either positive or negative, corresponding to an enhancement or suppression of the bending angle relative to the torsion-free case (GR). The qualitative behavior remains similar to that of the high-density white dwarf, with the axial-dominated configuration yielding the negative correction. However, due to the much lower fermion density, the correction is strongly suppressed and remains at the level of $\mathcal{O}(10^{-16})\,\mathrm{arcsec}$. In all cases, $|\alpha-\alpha_{\rm GR}|$ increases monotonically with $b/r_g$ and attains its maximum value near the stellar surface.

%
\begin{figure}[ht]
\includegraphics[width=0.5\textwidth]{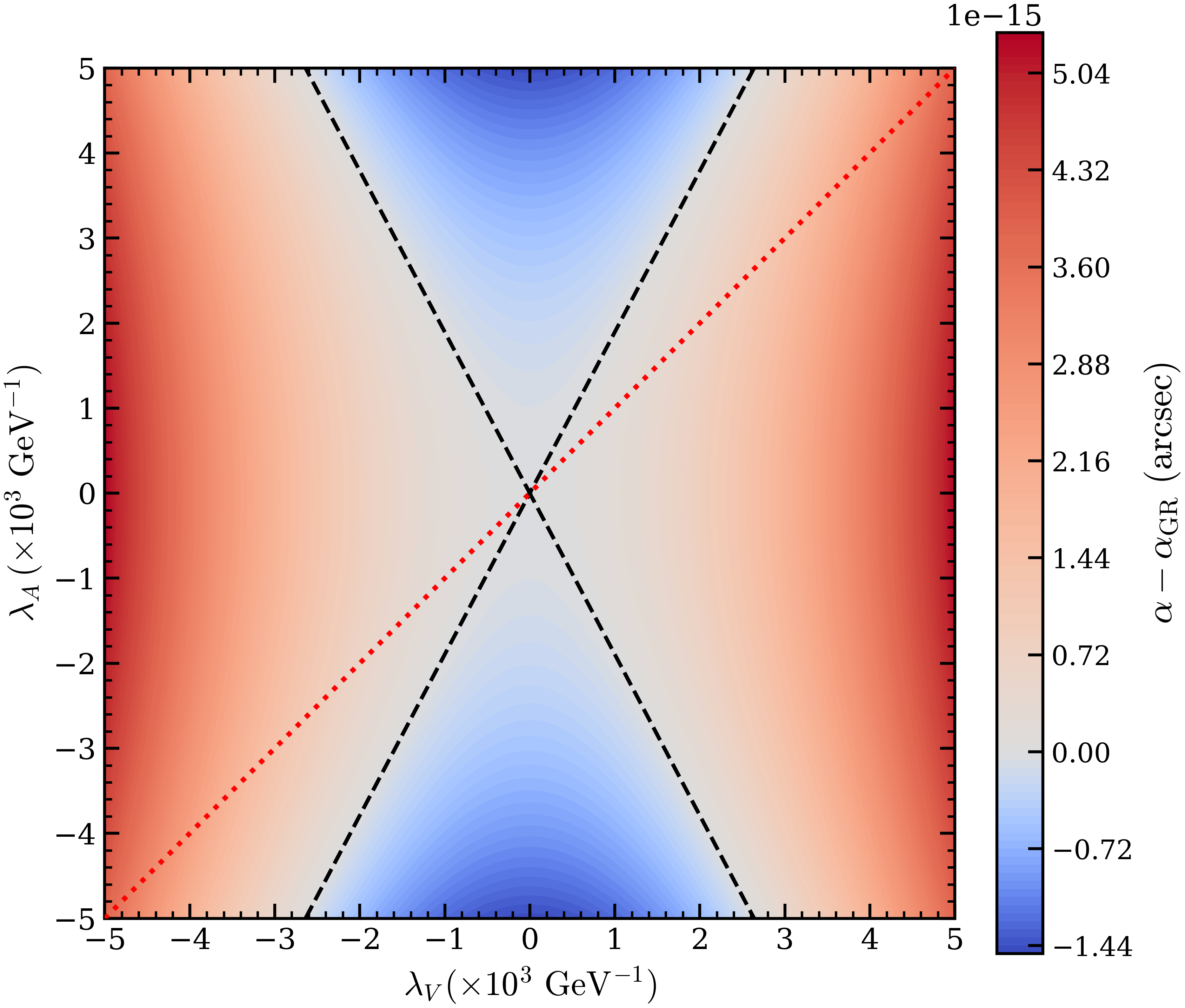}
\caption{
Filled contour plot of the torsion-induced correction to the bending angle $\alpha-\alpha_{\rm GR}$, in the $(\lambda_V, \lambda_A)$ parameter space for low density white dwarf with $b/r_g = 0.8$.  
}
\label{WD low mass_null_geodesic_bending contour plot} 
\end{figure}
%


Fig.~\ref{WD low mass_null_geodesic_bending contour plot} shows the correction $\alpha-\alpha_{\rm GR}$ for $b/r_g=0.8$, for a low-density white dwarf, in the $(\lambda_V,\lambda_A)$ parameter space. The plot has a similar pattern to that of the high-density white dwarf, with positive corrections in the vector-dominated regime (red regions) and negative corrections for sufficiently strong axial couplings (blue regions), occurring approximately for $|\lambda_A|\gtrsim 2.0\,|\lambda_V|$. The correction remains nonzero along the red dotted line, which represents the maximally chiral situation $\lambda_V=\lambda_A$, due to mixed fermionic interaction terms in Eq.~\eqref{bending angle for low density WD}. However, because of the substantially lower fermion density, the magnitude of the correction is even smaller than that of the highly dense white dwarf -- of the order $\mathcal{O}(10^{-15})\,\mathrm{arcsec}$.

We see that while torsion-induced corrections to null geodesic bending remain extremely small for both low- and high-density white dwarfs, their systematic dependence on fermion density, coupling strength, and chiral structure highlights the potential for significantly larger effects in more compact astrophysical objects, such as neutron stars, which we will explore in the following sections. 
\section{Null bending in neutron stars}\label{sec:results for NS} 
In this section we present a numerical study of null geodesic bending in the interior of a neutron star, modeled as a constant-density configuration. Since spin–torsion interactions can modify the effective gravitational mass of the star, it is convenient to characterize the stellar interior in terms of the baryon number density rather than the observed mass. The free part of the mass density is written in terms of a reference mass scale $M_0$ as $\rho_0=3M_0/(4\pi r_g^3)$, where $r_g$ denotes the stellar radius. The corresponding baryon number density is then $n_B=\rho_0/m_n$. For typical neutron-star core conditions, the baryon number density are expected to be a few times the nuclear saturation density \cite{Lattimer:2004pg, Lattimer:2000nx, Akmal:1998cf}. As a simplifying approximation, we assume a uniform interior matter distribution and take a constant baryon number density $n_B \simeq 3 n_0$, where $n_0 = 0.16\,\mathrm{fm}^{-3}$ is the nuclear saturation density, and a representative proton fraction $Y_p \simeq 0.05$. This yields neutron and proton number densities $n_n \simeq (1 - Y_p) n_B = 0.456\,\mathrm{fm}^{-3}, \quad n_p \simeq n_e \simeq Y_p n_B = 0.024\,\mathrm{fm}^{-3}$. The corresponding Fermi momenta are $k_{nF} \simeq 470\,\mathrm{MeV}, \quad k_{pF} \simeq 170\,\mathrm{MeV}, \quad k_{eF} \simeq 170\,\mathrm{MeV}$. These values indicate that the electrons are deeply ultra-relativistic $(z_e=k_{eF}/m_e \approx 330)$, whereas the nucleons remain non-relativistic, with $z_p=k_{pF}/m_p \simeq 0.2$ and $z_n=k_{nF}/m_n \simeq 0.5$. Therefore, for baryons, the term in expression in Eq.~\eqref{rhointresur}, $z_f \sqrt{1 + z_f^2 } -  \sinh^{-1} z_f \sim \frac{2z_f^3}{3} - \frac{z_f^5}{5} + \mathcal{O}(z_f^7)$, implying that the mass dependent contributions scale as $n_f^2$, and are therefore of the same parametric order as the density–density mean-field term $n_f^2/16$. However, in the ultra-relativistic limit for electrons, $z_e \sqrt{1 + z_e^2 } -  \sinh^{-1} z_e = z_e^2 - \ln(2z_e) +\frac{1}{2} - \frac{3}{8z_e^2} + \mathcal{O}(z_e^{-4})$, with the leading contribution proportional to $z_e^2$. {The corresponding mass-dependent term is thus suppressed by a factor $(m_e/k_{eF})^2 \ll 1$, makes it very small, compared to $n_f^2/16$}. The temperature dependent term is always very small for all species of fermions in a neutron star with temperature $10^8 \, K$. Consequently, Eq.~\eqref{rhointresur} reduces to  
\begin{equation}
    \rho^{\mathrm{int}} \simeq \frac{3}{16}\sum_{f \in (p, n)}(\lambda_{fV}^2-\lambda_{fA}^2) n_f^2 + \frac{1}{16}(\lambda_{eV}^2-\lambda_{eA}^2) n_e^2 + \sum_{\substack{f \neq f' \\ f,f' \in \{n,p,e\}}} \lambda_{fV}\lambda_{f'V}n_fn_{f'}. \label{rho int NS}
\end{equation}
We use this simplified expression to evaluate the effective mass \eqref{effective mass} and plot the bending angle given by Eq.~\eqref{total bending angle}. 
%
\begin{figure}[ht]
\centering
\includegraphics[width=0.45\textwidth]{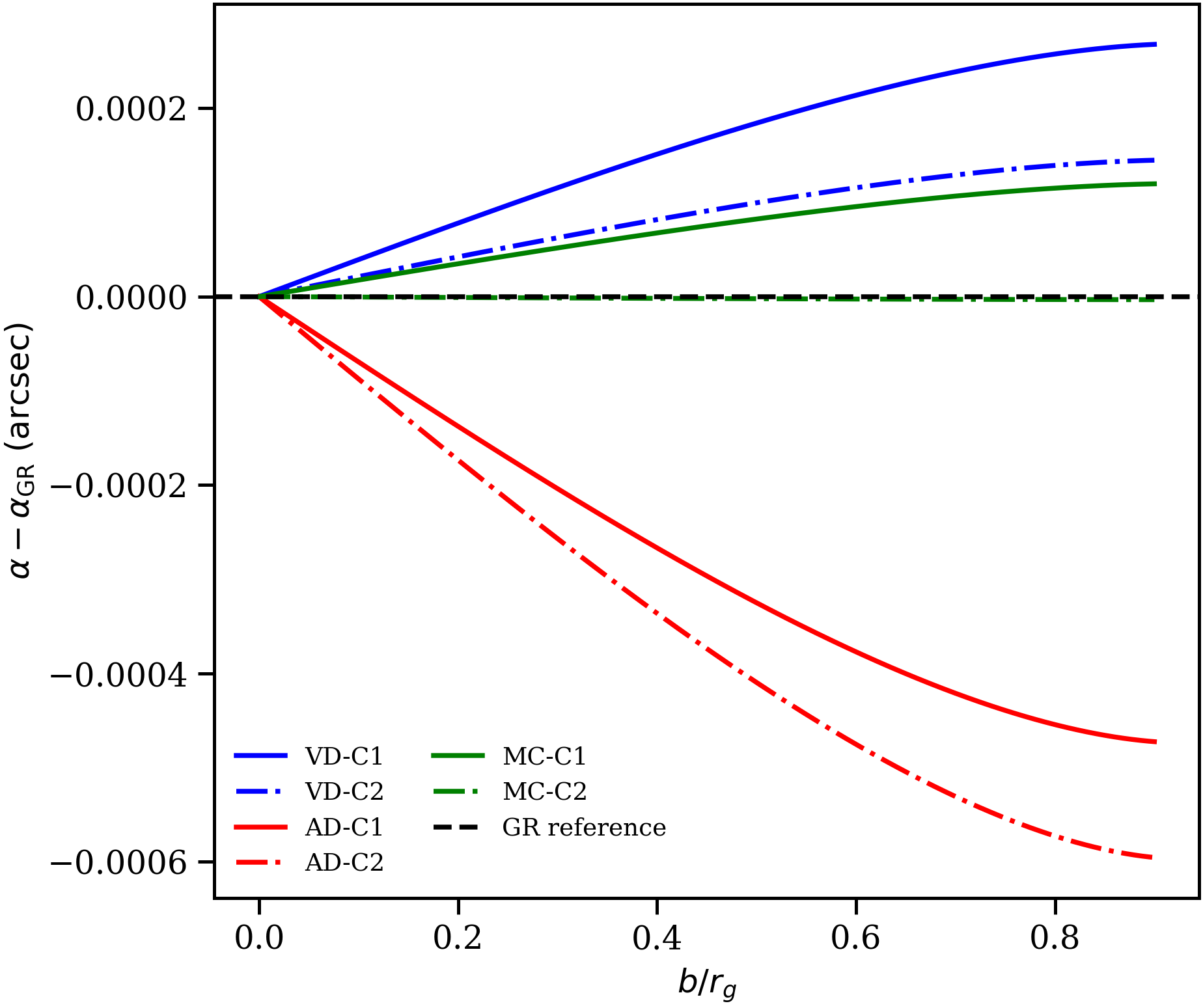}
\caption{Bending angle correction due to the spin-torsion interaction for Neutron star as a function of $b/r_g$ for different vector and axial coupling configurations.} 
\label{NS_null_geodesic_bending}
\end{figure}
In Fig.~\eqref{NS_null_geodesic_bending}, we show the behavior of the null–geodesic bending angle in the interior of a neutron star. The bending angle is evaluated using Eq. \eqref{total bending angle}, where the four-fermion interaction enters through the effective interaction energy density $\rho^{\mathrm{int}}$ given in Eq. \eqref{rho int NS}. The Fig.~\eqref{NS_null_geodesic_bending} shows the absolute correction to the bending angle, and thereby isolating the contribution arising solely from the interaction term. The correction can be either positive or negative depending on the values of the coupling constants $\lambda_V$ and $\lambda_A$. Its magnitude, $\lvert \alpha-\alpha_{\rm GR}\rvert$, increases monotonically with the normalized impact parameter $b/r_g$ and reaches its maximum near the stellar surface, similar to the behavior for white dwarfs. The bending angle correction is significantly enhanced in neutron stars due to their extremely high baryon number densities, reaching $\mathcal{O}(10^{-3}) \, \mathrm{arcsec}$ for the parameter values considered. In contrast, the much lower fermion densities in white dwarfs suppress the corresponding corrections by several orders of magnitude. These results indicate that neutron stars can be a sensitive probes of spin-induced interaction effects in gravitational lensing.

%
\begin{figure}[ht]
\includegraphics[width=0.5\textwidth]{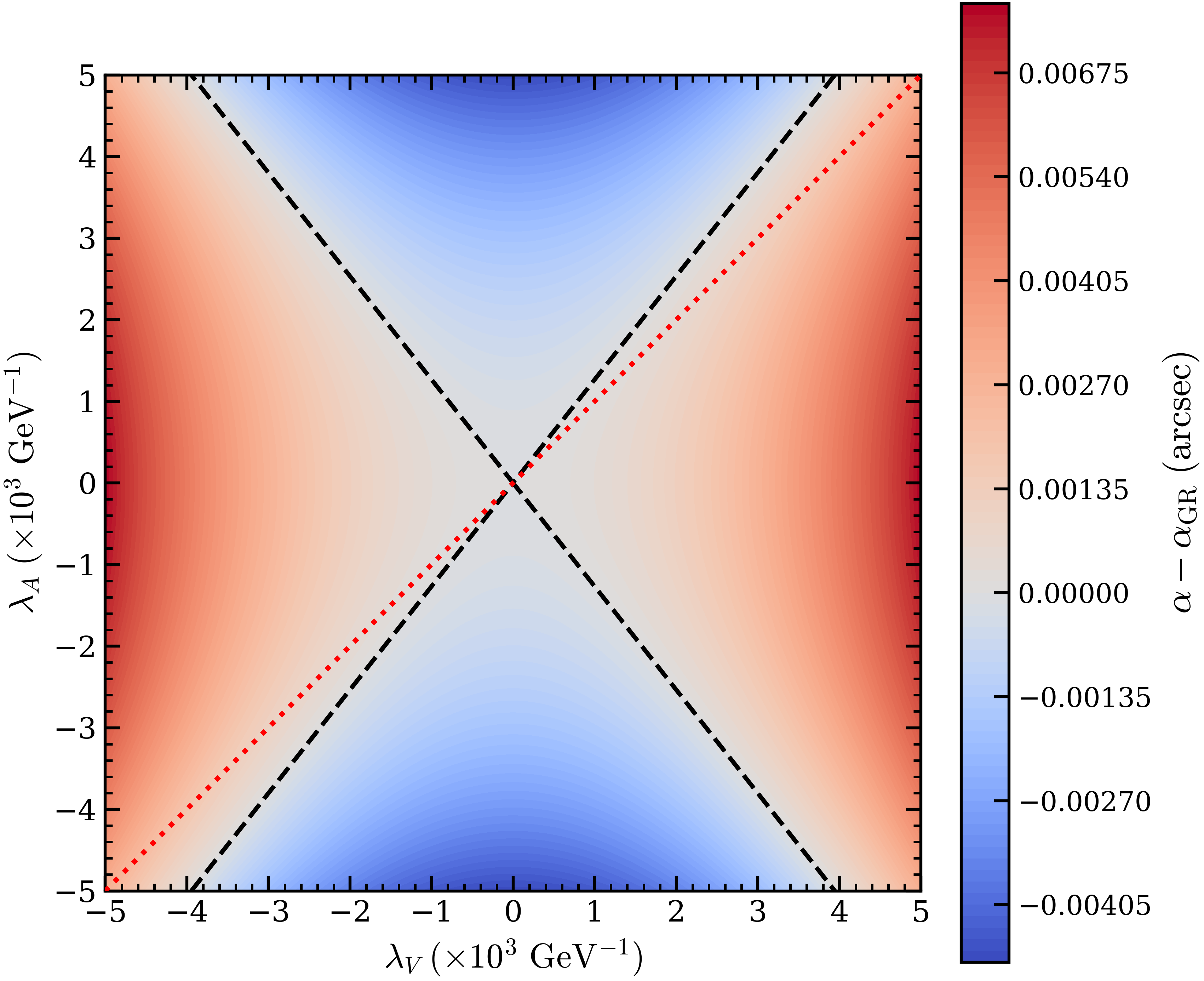}
\caption{
Filled contour plot of the torsion-induced correction to the bending angle $\alpha-\alpha_{\rm GR}$, in the $(\lambda_V, \lambda_A)$ parameter space for a neutron star, with $b/r_g = 0.8$.  
}
\label{NS_null_geodesic_bending contour plot} 
\end{figure}
%
 
We see that the plot for the case of the neutron star in Fig.~\eqref{NS_null_geodesic_bending contour plot} is quite similar to that of the white dwarf. There is a clear separation between regions of positive (red regions) and negative corrections (blue regions), corresponding to vector and axial dominated interactions, respectively. The bending angle correction $\alpha - \alpha_{\mathrm GR}$ becomes negative approximately when $|\lambda_A|\gtrsim 1.27\,|\lambda_V|$.
Since the neutron stars are predominantly composed of neutrons, the mixed fermionic contributions become strongly suppressed, and therefore, the red dotted line lies very close to the black dashed contour as shown in Fig.~\eqref{NS_null_geodesic_bending contour plot}. For the parameter range considered, the correction reaches $\mathcal{O}(10^{-3})\,{\rm arcsec}$, which is several orders of magnitude larger than for the white dwarf. For neutron stars, the torsion-induced correction is $\mathcal{O}(10^{-3}) \, {\rm arcsec}$, comparable to the precision level at which systematic effects become relevant in lensing-based mass determinations \cite{Kling:2007pw,Frittelli:2011uh}.

\section{Conclusion}\label{Conclusion}
In this work, we have studied the impact of fermion-induced torsion on geodesic bending inside compact objects such as white dwarfs and neutron stars. Within the Holst-modified ECSK framework, torsion is not purely antisymmetric and therefore contributes explicitly to the geodesic equations. The resulting spin–torsion interaction generates an effective four-fermion coupling, modifying the stellar energy density and consequently geodesic bending. For null geodesics, we derived the weak-field deflection angle and found that the torsion contribution arises through the modification of the effective energy density. For a single fermion species with a maximally chiral geometrical coupling, the energy density due to the spin-torsion interaction vanishes and the bending angle remains unchanged. We considered a general fermionic matter distribution composed of multiple species, each coupled chirally to the contorsion tensor through independent coupling constants. The correction to the bending angle vanishes at the center of the star and increases monotonically with the impact parameter and is maximum at the stellar surface. The torsional correction can enhance or reduce the bending angle depending on the sign and magnitude of the torsional couplings and the density and temperature of the background fermions.

{We have computed the change in the deflection angle for realistic white dwarf and neutron star configurations, by considering only gravity and torsion and neglecting all other interactions. We found corrections of order $10^{-3}$ arcsec for neutron stars, $10^{-13}$ arcsec for high-density white dwarfs, and $10^{-16}$ arcsec for low-density white dwarfs. This hierarchy reflects the strong dependence of the spin--torsion contribution on the fermion number density, with the much larger densities in neutron stars producing substantially larger corrections than in white dwarfs. These results illustrate how geodesic bending can, in principle, carry information about microscopic spin-torsion interactions in dense fermionic matter.}

Diverging gravitational lensing can arise in a variety of astrophysical and cosmological settings, including systems with negative effective energy densities~\cite{Kitamura:2012zy, Izumi:2013tya, PhysRevD.90.084026, Er:2017lue} and underdense cosmic voids that act as weak defocusing lenses~\cite{Piran:1997fs, Amendola:1998xu, Melchior:2013gxd}. In our work, regions of parameter space in which axial couplings dominate can produce negative corrections to the effective energy density.  These effects could become more pronounced in compact objects with very high fermion densities, where the spin–torsion contribution may be substantially enhanced. Exploring such scenarios requires extending our analysis to realistic stellar density profiles obtained from the Tolman–Oppenheimer–Volkoff equations. An interesting application of our method will be on the impact of spin–torsion interactions on the mass-radius relation in compact stars, including quark stars and stars containing dark-matter~\cite{Narain:2006kx, Kouvaris:2015rea, Ellis:2018bkr, Mukhopadhyay:2015xhs, Barbat:2024yvi}.   

Another situation where the spin-torsion interaction is expected to be significant is the early universe close to the initial singularity~\cite{Dolan:2009ni, Bambi:2014uua, Alexander:2008vt}. At the densities and energies prevalent at that time, one expects gauge couplings to be negligible and the torsional four-fermion interaction could become dominant. A related scenario is that of the propagation of relic neutrinos in cosmological backgrounds, which has attracted considerable attention recently~\cite{Lin:2019lko, Baym:2020riw, Baym:2021ksj, Taak:2022xdp, Hernandez-Molinero:2023jes, Ge:2024kac}. It would be interesting to study whether spin-torsion interactions can similarly modify the propagation of relic neutrinos, and whether such effects can leave observable signatures in cosmological lensing phenomena. We have considered the torsion induced interactions only in the background fermionic matter but ignored such interactions between the background and a test fermion traveling through it. Recently some papers have investigated how to derive a geodesic for a spin-$\frac{1}{2}$ particle from the Dirac equation on a curved background~\cite{Hammad:2024dqp, Hammad:2024dhf, Oancea:2022utx, Cianfrani:2008hq, rudiger1981dirac, audretsch1981trajectories}, and the resulting particle deflection~\cite{Li:2025mcp, Zhang:2022rnn}. It will be interesting to apply their methods to a particle that also interacts torsionally with the background.

\appendix
\section{Energy density and Number density of degenerate fermion matter}\label{appendix:a}
We consider a more general integration, 
\begin{equation}
    I = \int_m^\infty g(\epsilon) \frac{d\epsilon}{e^{\beta(\epsilon - \mu)}+1}. \label{generic integral}
\end{equation}
For $\beta(\mu - m) \gg 1$ we have~\cite{sommerfeld1928elektronentheorie, fukushima2014analytical}  
\begin{equation}
    I = \int_m^\mu g(\epsilon) d\epsilon + 2\sum_{k=0}^\infty \frac{g^{(2k+1)}(\mu)}{\beta^{2k+2}} \big( 1 - 2^{-2k-1}\big) \zeta(2k+2)\,, \label{lowtemp}
\end{equation} 
where $\zeta$ is the Riemann zeta function and   $g^{(2k+1)}(\mu)
= \left.\frac{d^{\,2k+1} g(\epsilon)}{d\epsilon^{\,2k+1}}\right|_{\epsilon=\mu}$. 
Taking only the {$k=0$} term, we can write 
\begin{equation}
    I = \int_{m}^\mu g(\epsilon) d\epsilon + \frac{\pi^2}{6} \big( k_B T \big)^2 g'(\mu) + \mathcal{O}\big( T^4 \big) \,. \label{Sommerfeld_generic}
\end{equation}
Applying Eq.~\eqref{Sommerfeld_generic} with {$g(\epsilon)=\mathcal{D}(\epsilon)$} and retaining terms up to order $T^2$, we write the number density \eqref{numberdensity} as 
\begin{equation}
    n = \int_{m}^\mu \mathcal{D}(\epsilon) d\epsilon + \frac{\pi^2}{6} (k_B T)^2 \mathcal{D}'(\mu) + \mathcal{O}(T^4)\,.
\end{equation}
We now look at the free part of the energy density. The general expression for the free part of the energy is given in Eq.~\eqref{rho free part}. This is the generic form of Eq.~\eqref{generic integral} with $g(\epsilon) = \epsilon \mathcal{D}(\epsilon)$. Thus from Eq.~\eqref{Sommerfeld_generic}, we can write
\begin{equation}
    \rho^{\text{free}} =  \int_{m}^\mu \epsilon\mathcal{D}(\epsilon) d\epsilon + \frac{\pi^2}{6} (k_B T)^2 \Big[\dv{}{\epsilon}\epsilon\mathcal{D}(\epsilon)\Big]_{\epsilon=\mu} + \mathcal{O}(T^4)\,.
\end{equation}
The chemical potential $\mu$ \eqref{muexpression2} at $T=0$ is the Fermi energy $\epsilon_F$, so the energy density at $T=0$ is given by
\begin{equation}
    \rho^{\text{free}}_0 = \int_{m}^{\epsilon_F} \epsilon\mathcal{D}(\epsilon) d\epsilon\,.
\end{equation}
Thus
\begin{equation}
    \rho^{\text{free}} = \rho^{\text{free}}_0 + \int_{\epsilon_F}^\mu \epsilon\mathcal{D}(\epsilon) d\epsilon + \frac{\pi^2}{6} (k_B T)^2 \big[ \mathcal{D}(\mu) + \mu \mathcal{D}'(\mu)\big] + \mathcal{O}(T^4)\,.
\end{equation}
Since the temperature is very low, we can assume that $\mu$ is close to $\epsilon_F$. In the remaining explicit integral, we can then assume the integrand to be a constant, equal to the value of the integrand at $\epsilon_F$. Also in the $\mathcal{O}(T^2)$ term, we can replace the density of states at $\epsilon = \mu$ by the corresponding quantity at $\epsilon = \epsilon_F$, since the difference would contribute at higher order in $T$. So finally
\begin{equation}
    \rho^{\text{free}} = \rho^{\text{free}}_0 + (\mu - \epsilon_F) \epsilon_F \mathcal{D}(\epsilon_F) + \frac{\pi^2}{6} (k_B T)^2 \big[ \mathcal{D}(\epsilon_F) + \epsilon_F \mathcal{D}'(\epsilon_F)\big] + \mathcal{O}(T^4)\,.
\end{equation}
Therefore using Eq.~\eqref{muexpression2}, we obtain the final expression as
\begin{equation}
    \rho^{\text{free}} = \rho^{\text{free}}_0 + \frac{\pi^2}{6} (k_B T)^2 \sum_f \mathcal{D}(\epsilon_{fF}) + \mathcal{O}(T^4)\,,  
\end{equation}

In the following we calculate the energy density for the interaction part in the approximation $\beta(\mu - m) \gg 1\,.$ We expand the following integral using Eq.~\eqref{Sommerfeld_generic} with $g(\epsilon) = \sqrt{\epsilon^2 - m^2 }$ as  
\begin{equation}
    \int d\epsilon \frac{\sqrt{\epsilon^2 - m^2 }}{e^{\beta(\epsilon - \mu)}+1} = \frac{1}{2} \Big[ \epsilon_F \sqrt{\epsilon_F^2 - m^2 } - m^2  \sinh^{-1}\Big(\frac{k_F}{m}\Big) \Big] + \frac{\pi^2}{6} \big(k_B T\big)^2 \frac{\sqrt{\epsilon_F^2 - m^2 }}{\epsilon_F} + \mathcal{O}(T^4)\,.
\end{equation}
The expectation value of the temporal component of the vector current is
\begin{equation}
    \langle J_{V0}\rangle = \frac{1}{\pi^2} \int d\epsilon \frac{\epsilon \sqrt{\epsilon^2 - m^2 }}{e^{\beta(\epsilon - \mu)}+1} = n\,. \label{expectation JV0}
\end{equation}
Therefore, in small temperature approximation Eq.~\eqref{expectation JVI JVI} simplifies to
\begin{align} \label{expectation JVI JVI final}
    \langle J_{Va} J^{Va} \rangle &= - \frac{m^2}{4\pi^4} \Bigg[ \frac{1}{4} \Bigg(\epsilon_F \sqrt{\epsilon_F^2 - m^2 } - m^2  \sinh^{-1} \big(\frac{k_F}{m}\big) \Bigg)^2 \\ \notag
    &+ \frac{\pi^2}{6} (k_B T)^2 \frac{\sqrt{\epsilon_F^2 - m^2 }}{\epsilon_F} \Bigg(\epsilon_F \sqrt{\epsilon_F^2 - m^2 } - m^2  \sinh^{-1} \big(\frac{k_F}{m}\big) \Bigg)\Bigg] - \frac{1}{8} \Bigg[ \frac{k_F^3}{3\pi^2} \Bigg]^2\,.
\end{align}
We use Eq.~\eqref{expectation JVI JVI final} and \eqref{expectation JV0} in Eq.~\eqref{rhointeraction} and get the interaction part of the energy density calculated in Eq.~\eqref{rhointresur}.
\section{Geodesic equations in the presence of fermionic matter}\label{appendix:b}
We restrict our consideration of geodesics to the equatorial plane ($\theta = \pi/2$) and therefore the geodesic equations for $\phi$ and $t$ coordinates becomes,
\begin{equation}
    \ddot{\phi} + \frac{2}{r} \dot{\phi} \dot{r} = - \xi \Big[ \big(u^\rho J_\rho \big) \dot{\phi} - \epsilon \frac{1}{r} J_3 \Big]. \label{nulltimelikephiequ}
\end{equation}
\begin{equation}
    \ddot{t} + \frac{B'(r)}{B(r)} \dot{t} \dot{r} = - \xi \Big[ \big(u^\rho J_\rho \big) \dot{t} + \epsilon \frac{1}{\sqrt{B(r)}} J_0 \Big]. \label{nulltimeliketequ}
\end{equation}
For timelike $\epsilon = -1$, and for null geodesics, it is 0. The term ($u^\rho J_\rho$) is constant along the geodesics \eqref{urhojrho}, therefore we write \eqref{nulltimelikephiequ} as
\begin{equation}
    \dv{}{\lambda} \dv{\phi}{\lambda} + \Bigg[\frac{2}{r} \dv{r}{\lambda} + \xi a \Bigg]\dv{\phi}{\lambda} = \epsilon \xi \frac{1}{r} J_3 . \label{timelikephiequ2}
\end{equation}
Let's define $x = \dv{\phi}{\lambda}$, therefore, we write \eqref{timelikephiequ2} as the first order differential equation,
\begin{equation}
    \dv{x}{\lambda} + P(\lambda) x = Q(\lambda). \label{timelikephiequ3}
\end{equation}
which can be solve easily by defining the integrating factor $\mu = \exp \Big( \int d\lambda P(\lambda) \Big)$,
\begin{equation}
    \mu = \exp \Big( \int dr \frac{2}{r} + \int d\lambda \xi a \Big) = C_1 r^2 e^{\xi a \lambda}
\end{equation}
We now multiplying the integrating factor both side of \eqref{timelikephiequ3}, and we get, 
\begin{equation}
    \dv{}{\lambda} \Big[ r^2 e^{\xi a \lambda} x \Big] = r^2 e^{\xi a \lambda} Q(\lambda)
\end{equation}
Hence, 
\begin{equation}
    r^2 \dv{\phi}{\lambda} = L e^{-\xi a \lambda} + \epsilon \xi J_3 e^{-\xi a \lambda} \int_0^{\lambda(r)} d\lambda' r e^{\xi a \lambda'}  
\end{equation}
Similarly, we can write \eqref{nulltimeliketequ} as
\begin{equation}
    B(r) \dv{t}{\lambda} = E e^{-\xi a \lambda} - \epsilon \xi J_0 e^{-\xi a \lambda} \int_0^{\lambda(r)} d\lambda' B^{\frac{1}{2}}(r) e^{\xi a \lambda'}
\end{equation}
For null geodesic ($\epsilon = 0$)
we get \eqref{nullphitequ}, and for timelike geodesic ($\epsilon = -1$, the affine parameter is the proper time $\tau$) we get the equations \eqref{timelikephitequ}.
\bibliographystyle{unsrt}
\bibliography{Bibliography}
 
\end{document}